\documentclass[acmsmall]{acmart}
\AtBeginDocument{%
  }

\usepackage{graphicx}
\usepackage{textcomp}
\usepackage{xcolor}
\usepackage[utf8]{inputenc}
\usepackage[T1]{fontenc}
\usepackage{url}
\usepackage{booktabs}
\usepackage{tabularx}
\usepackage{nicefrac}
\usepackage{microtype}
\usepackage{xcolor}
\usepackage{multirow}
\usepackage{graphicx}
\usepackage{comment}
\usepackage{array}
\usepackage{colortbl}
\usepackage{subcaption}
\usepackage{caption}
\usepackage{enumitem}
\usepackage{pifont}
\usepackage[most]{tcolorbox}
\usepackage{wrapfig}
\usepackage{float}
\usepackage{enumitem}
\usepackage{fontawesome}
\usepackage{tikz}
\usepackage{listings}
\usepackage{longtable}

\setcopyright{none}
\renewcommand{\footnotetextcopyrightpermission}[1]{}
\begin{document}

\newcommand{\bench}[0]{\textsc{CRBugBench}}
\newcommand\circleone{\ding{192}}
\newcommand\circletwo{\ding{193}}
\newcommand\circlethree{\ding{194}}
\newcommand\circlefour{\ding{195}}
\newcommand\circlefive{\ding{196}}
\newcommand\circlesix{\ding{197}}
\newcommand\circleseven{\ding{198}}
\newcommand\circleeight{\ding{199}}

\definecolor{mybgcolor}{HTML}{E6F2F0}    %
\definecolor{myframecolor}{HTML}{FF8552} %
\definecolor{mytitlecolor}{HTML}{2C7873} %

\definecolor{tblheader}{HTML}{F5E6DC}   %
\definecolor{tblheadtext}{HTML}{5B3A1E}  %
\definecolor{adaptgray}{HTML}{999999}    %

\newcommand{\nbc}[3]{
 {\colorbox{#3}{\bfseries\sffamily\scriptsize\textcolor{white}{#1}}}
 {\textcolor{#3}{\sf\footnotesize$\blacktriangleright$\textit{#2}$\blacktriangleleft$}}
 }

\newcommand\todo[1]{\nbc{TODO}{#1}{red}}
\newcommand{\cyang}[1]{\nbc{CY}{#1}{teal}}
\newcommand{\swcomment}[1]{\nbc{SW}{#1}{blue}}
\newcommand{\ck}[1]{\nbc{CK}{#1}{orange2}}
\newcommand{\emh}[1]{\nbc{EM}{#1}{violet}}
\newcommand{\cma}[1]{\nbc{CM}{#1}{orange}}

\definecolor{boxcolor}{RGB}{238, 223, 204} %
\DeclareRobustCommand{\mybox}[2][gray!20]{%
\begin{tcolorbox}[   %
        breakable,
        left=0pt,
        right=0pt,
        top=0pt,
        bottom=0pt,
        colback=#1,
        colframe=black,
        width=\dimexpr\columnwidth\relax, 
        enlarge left by=0mm,
        boxsep=5pt,
        outer arc=4pt,
        boxrule=.5mm
        ]
        #2
\end{tcolorbox}
}

\newcommand{\delete}[1]{\textcolor{deeporange}{\st{#1}}}
\newcommand{\new}[1]{{#1}}

\newcounter{hyp}
\newcommand{\Hypothesis}[1]{%
  \refstepcounter{hyp}%
  \textbf{Hypothesis~\thehyp}: {#1}\label{hyp:\thehyp}%
}

\newcounter{rq}
\newcommand{\RQ}[1]{%
  \refstepcounter{rq}%
  \textbf{RQ~\therq}: {#1}\label{rq:\therq}%
}

\definecolor{tagbg}{RGB}{245,237,255}
\definecolor{tagtext}{RGB}{110,72,170}

\DeclareRobustCommand{\dimtag}[1]{%
  \tcbox[
    on line,
    nobeforeafter,
    tcbox raise base,
    colback=tagbg,
    colframe=tagbg,
    arc=3pt,
    boxsep=0pt,
    left=3pt,
    right=3pt,
    top=1pt,
    bottom=1pt,
    boxrule=0pt,
    fontupper=\sffamily\bfseries\footnotesize\color{tagtext},
    before upper={\strut}
  ]{#1}%
}

\DeclareRobustCommand{\tabledimtag}[1]{%
  \tcbox[
    on line,
    nobeforeafter,
    tcbox raise base,
    colback=tagbg,
    colframe=tagbg,
    arc=2pt,
    boxsep=0pt,
    left=2pt,
    right=2pt,
    top=0pt,
    bottom=0pt,
    boxrule=0pt,
    fontupper=\sffamily\bfseries\scriptsize\color{tagtext}
  ]{#1}%
}

\lstset{
  language=Python,
  keywords={},
  commentstyle=\textbf,
  stringstyle=\text,
  extendedchars=false,
  basicstyle=\ttfamily,
  columns=fullflexible,
  frame=single,
  breaklines=true,
  breakatwhitespace=true,
  breakindent=0\dimen0,
  literate={'}{{\textquotesingle}}1 {`}{{\textasciigrave}}1 {"}{{\textquotedbl}}1,
}

\newcommand{\MinedPrs}{304{,}601}
\newcommand{\MinedThreads}{1{,}436{,}120}
\newcommand{\RetainedThreads}{472{,}039}
\newcommand{\MinedFixPrs}{195{,}075}
\newcommand{\MinedFixPrsWithIssue}{57{,}725}
\newcommand{\BenchTasks}{1{,}000}
\newcommand{\BenchRepos}{98}
\newcommand{\MedianChurn}{189}
\newcommand{\SingleLanguageShare}{82}
\newcommand{\BestModel}{\texttt{claude-opus-5}}
\newcommand{\BestCmtRecall}{79.6}
\newcommand{\BestFixRecall}{64.2}
\newcommand{\BestRecall}{71.9}
\newcommand{\BestCost}{2.97}
\newcommand{\BestLatency}{9.8}
\newcommand{\EvolRecallOld}{47.7}
\newcommand{\EvolRecallNew}{71.9}
\newcommand{\EvolCmtOld}{59.1}
\newcommand{\EvolCmtNew}{79.6}
\newcommand{\SmallModel}{\texttt{gpt-5.6-luna}}
\newcommand{\SmallRecall}{55.9}
\newcommand{\SmallCost}{0.62}
\newcommand{\SmallRecallShare}{78}
\newcommand{\SmallCostShare}{21}
\newcommand{\PromptGain}{5.8}
\newcommand{\PromptCostRatio}{1.10}
\newcommand{\PromptGainMean}{3.8}
\newcommand{\PromptCostRatioMean}{0.99}
\newcommand{\PromptGainSol}{5.3}
\newcommand{\EffortModel}{\texttt{gpt-5.6-luna}}
\newcommand{\EffortLowRecall}{36}
\newcommand{\EffortMaxRecall}{57}
\newcommand{\EffortCostRatio}{23.22}
\newcommand{\MultiAgentGain}{4.7}
\newcommand{\MultiAgentCostRatio}{5.55}
\newcommand{\MultiAgentGainMean}{2.8}
\newcommand{\MultiAgentCostRatioMean}{3.68}
\newcommand{\SubagentsBest}{9.0}
\newcommand{\SubagentsPrior}{0.1}
\newcommand{\SubagentsCopilot}{0}
\newcommand{\HumanRecall}{50}
\newcommand{\AgentRecall}{71.9}
\newcommand{\UnionRecall}{82.1}
\newcommand{\AdoDrop}{25.5}
\newcommand{\AdoVisibleShare}{22.8}
\newcommand{\GithubVisibleShare}{46.1}
\newcommand{\AdoRateAtGithubVisibility}{46.9}
\newcommand{\AdoRate}{42.1}
\newcommand{\GithubRate}{63.0}
\newcommand{\AdoGapOr}{0.407}
\newcommand{\AdoGapOrLocality}{0.488}
\newcommand{\AdoLocalityExplained}{20}
\newcommand{\HitTopThree}{89}
\newcommand{\OrDiffSize}{0.591}
\newcommand{\SigDiffSize}{***}
\newcommand{\PDiffSize}{$<$0.001}
\newcommand{\OrHeterogeneity}{0.795}
\newcommand{\SigHeterogeneity}{*}
\newcommand{\PHeterogeneity}{0.01}
\newcommand{\OrVisible}{2.621}
\newcommand{\SigVisible}{***}
\newcommand{\PVisible}{$<$0.001}
\newcommand{\OrSeverity}{1.906}
\newcommand{\SigSeverity}{**}
\newcommand{\PSeverity}{0.002}
\newcommand{\OrBugfixPR}{0.521}
\newcommand{\SigBugfixPR}{***}
\newcommand{\PBugfixPR}{$<$0.001}
\newcommand{\OrUsability}{0.454}
\newcommand{\SigUsability}{***}
\newcommand{\PUsability}{$<$0.001}
\newcommand{\OrPortability}{0.523}
\newcommand{\SigPortability}{n.s.}
\newcommand{\PPortability}{0.07}
\newcommand{\MaxRegression}{8.9}
\newcommand{\BestRegression}{1.2}
\newcommand{\OrTopOneRefactor}{0.568}
\newcommand{\SigTopOneRefactor}{*}
\newcommand{\OrTopOneOther}{0.476}
\newcommand{\SigTopOneOther}{***}
\newcommand{\OrHumanMaintainability}{2.702}
\newcommand{\SigHumanMaintainability}{*}
\newcommand{\OrHumanReliability}{0.591}
\newcommand{\SigHumanReliability}{*}
\newcommand{\OrHumanSeverity}{0.450}
\newcommand{\SigHumanSeverity}{*}
\newcommand{\ActionRead}{39}
\newcommand{\ActionProbe}{36}
\newcommand{\ActionExecute}{8}
\newcommand{\ActionWrite}{6}
\newcommand{\ActionGit}{11}
\newcommand{\ActionInspect}{75}
\newcommand{\FirstTurnGit}{64}
\newcommand{\ExecuteLong}{9.7}
\newcommand{\HalfLifeLong}{35}
\newcommand{\SurviveLong}{25.7}
\newcommand{\ExecuteShort}{6.9}
\newcommand{\HalfLifeShort}{17}
\newcommand{\SurviveShort}{3.5}
\newcommand{\GateInspect}{95.6}
\newcommand{\GateReport}{80.2}
\newcommand{\GateDiagnose}{56.2}
\newcommand{\BestGateReport}{91.9}
\newcommand{\BestGateDiagnose}{75}

\newcommand{\CB}[1]{\textcolor{red}{[CB: #1]}}

\title{(When) Should a Software Factory be Lights-off? Benchmarking Code Review Agents on Bug Detection}

\author{Chenyang Yang}
\authornote{Work done at Microsoft Research.}
\affiliation{%
  \institution{Carnegie Mellon University}
  \country{USA}
}

\author{Hao He}
\author{Ti-Chung Cheng}
\affiliation{%
  \institution{Microsoft}
  \country{USA}
}

\author{Jenna Butler}
\author{Emerson Murphy-Hill}
\affiliation{%
  \institution{Microsoft}
  \country{USA}
}

\author{Carmen Badea}
\author{Christian Bird}
\affiliation{%
  \institution{Microsoft Research}
  \country{USA}
}

\renewcommand{\shortauthors}{Yang et al.}

\begin{abstract}
With the advances of coding agents, software developers are spending more time \textit{reviewing} code than writing code.
As human review velocity falls behind agentic code generation, it is tempting to build automated review agents towards a future of ``lights-off'' software factory with little to no human supervision. 
To assess the practicality of such a fully autonomous software factory, we construct \bench{}, a new benchmark that evaluates whether review agents detect human-verified bugs.
We build a data pipeline that mines bugs from repository history, and pair each bug with its introducing PR to form a review task. 
\bench{} consists of \BenchTasks{} review tasks across \BenchRepos{} open-source GitHub repositories, and 800 additional tasks from proprietary industry repositories at one large company.

We then use the benchmark to study agent review quality: We found that SOTA review agents do well: the best (\BestModel{} with \texttt{copilot-cli}) catches \EvolRecallNew\% of bugs within its top-5 comments, up from \EvolRecallOld\% (\texttt{opus-4.6}) in five months of model evolution.
Yet agents remain less effective on complex changes and bugs requiring non-local reasoning, with sharp performance drops on proprietary industry repositories where project context is more often needed.
To improve bug detection, we found scaling compute helps most through \emph{prompt} changes: exploration-nudging prompts gain up to \PromptGain\% performance at 10\% higher cost.
From these findings, we derive concrete guidance for developers to adopt review agents -- scaling review effort to PR complexity, steering agents by project priorities, supplying repository knowledge, dispatching models by PR characteristics, and monitoring quality as models evolve.
\end{abstract}

\received{20 February 2007}
\received[revised]{12 March 2009}
\received[accepted]{5 June 2009}

\maketitle

\section{Introduction}
Software engineering is increasingly automated.
From code completion~\cite{bird2022taking}, chat assistants~\cite{liang2024large}, to coding agents~\cite{anthropicClaudeCode}, 
the task of writing code has been shifted from human developers to AI.
Developers are now increasingly doing more \textit{verification} work, commonly through code review~\cite{agarwal20263100}.
Yet the velocity of human reviews cannot catch up with agent-generated code.
It is tempting to further automate humans away, by using AI for code reviews~\cite{anthropicCodeReview}.
The extreme version of this would be a so-called ``lights-off'' software factory~\cite{shapiro2026fivelevels}, where AI both writes and reviews code, with little to no human supervision in the loop.

Are current repositories ready to become lights-off repositories?
To answer that question, we first need to understand how well AI can \textit{review} code.
Without good review capability, the engineering loop cannot be fully closed.
While many code review benchmarks exist (e.g., CodeReviewer~\cite{li2022automating}, SWR-Bench~\cite{zeng2025benchmarking}, c-CRAB~\cite{zhang2026code}),
they typically treat human comments as ground truth and expect AI-generated reviews to match them, lexically or semantically. 
Human comments, without filtering, are a poor proxy of review ground truth.
They can be arbitrary, containing nitpicking or less useful comments that are accepted due to social dynamics~\cite{bosu2015characteristics, sarma2024effective}.
They can also be too narrow, missing valid points that AI reviews can bring up.
Therefore, assessing review quality based on the ground truth of human reviews is misleading:
AI may provide high-quality review, just differently than human reviewers~\cite{tantithamthavorn2026rovodev}.
It would be hard to cleanly interpret the outcomes from these benchmarks.

\begin{figure}[t]
    \centering
    \includegraphics[width=\linewidth]{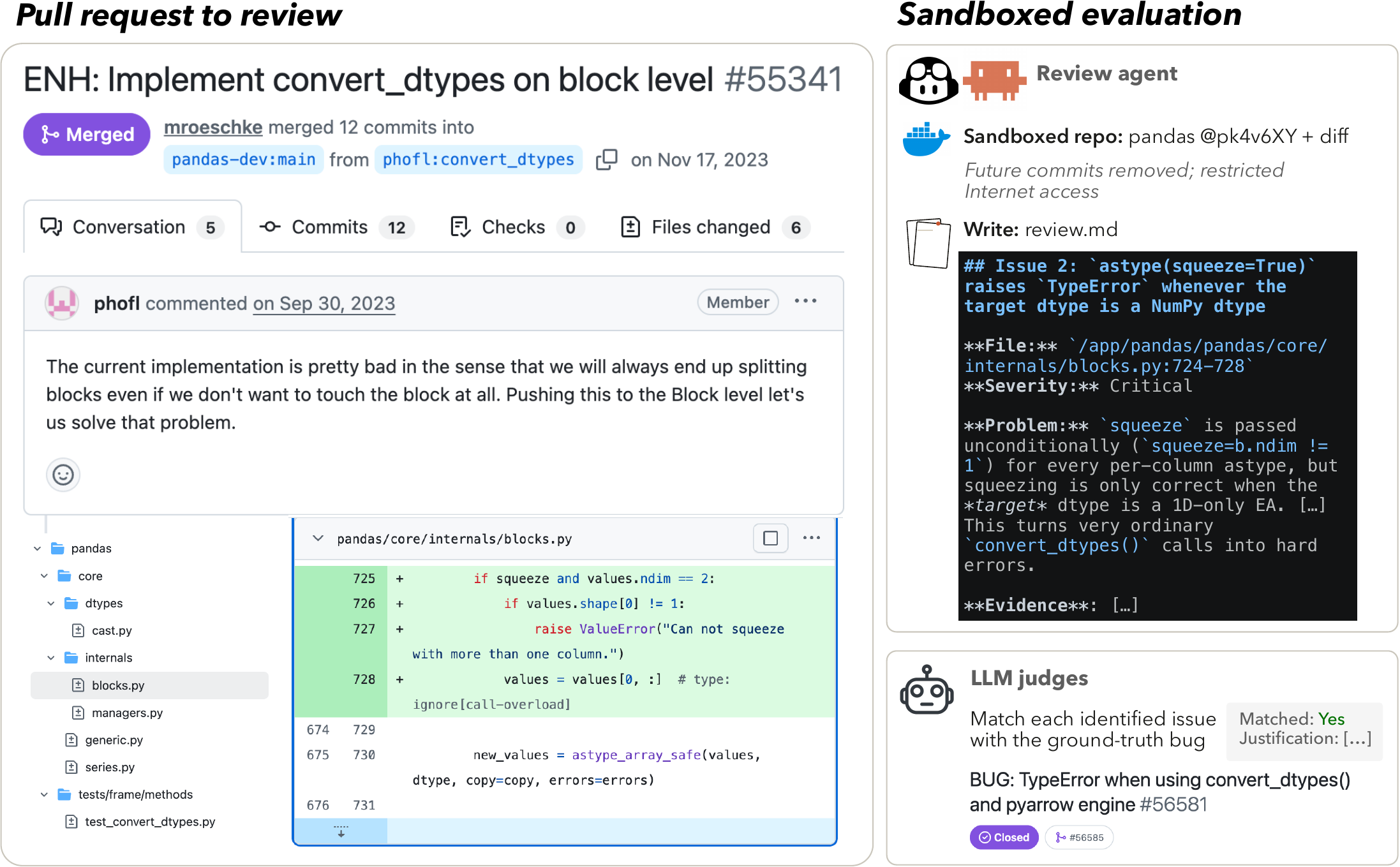}
    \caption{\bench{} identifies historical human-verified bugs, and traces them back to the bug-inducing PR commit. Review agents are provided with a sandboxed repository at the commit to explore and experiment, producing a final \texttt{review.md} file, with findings self-ranked by severity.
    A panel of LLM judges then compares each finding in the review to the ground-truth bug, and labels whether there are matches.
    }
    \vspace{-15pt}
    \label{fig:task_example}
\end{figure}

To provide a more objective assessment of review quality, 
we \textbf{curated a code review benchmark that evaluates how well a PR review detects human-verified bugs}.
The intuition is that, a good PR review, should at least catch the bugs that a human has confirmed and fixed.
We design, iterate, and validate a scalable data pipeline to collect diverse bugs from repository history.
We applied this data pipeline to both open-source repositories on GitHub and industry proprietary code repositories at one large company and collected verified bugs from two sources: 
First, from addressed human comments in PR discussions -- these are bugs that are identified and fixed \textit{before} the PR is merged. 
We identified them by finding all human review threads and filtering them down to the ones that leave bug-related comments and are addressed later.
Second, from human-submitted bug-fix PRs -- these are bugs that are identified \textit{after} a PR is merged.
We identified them by finding all bug-fixing issues and PRs, and traced them back to the PRs that introduced the bugs.
These bugs then serve as the ground truth that a review agent should identify from the introducing PRs.

In the end, we curated \bench, a bug-grounded code review benchmark with 1{,}800 code review tasks, across \BenchRepos{} GitHub open-source repositories, 2 large industry repositories, and 5 programming languages.
These tasks cover diverse PR sizes and types, as well as diverse bugs with different categories, sources, and complexity (Figure~\ref{fig:repo-stats}).

\begin{figure}[t]
    \centering
    \begin{subfigure}[t]{0.30\linewidth}
        \centering
        \includegraphics[width=\linewidth]{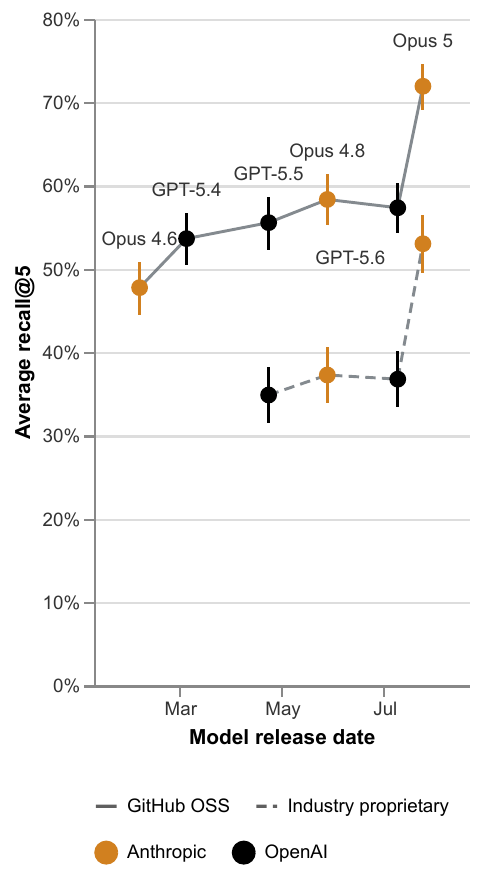}
        \caption{}
        \label{fig:performance-time-trend}
    \end{subfigure}
    \hfill
    \begin{subfigure}[t]{0.67\linewidth}
        \centering
        \includegraphics[width=\linewidth]{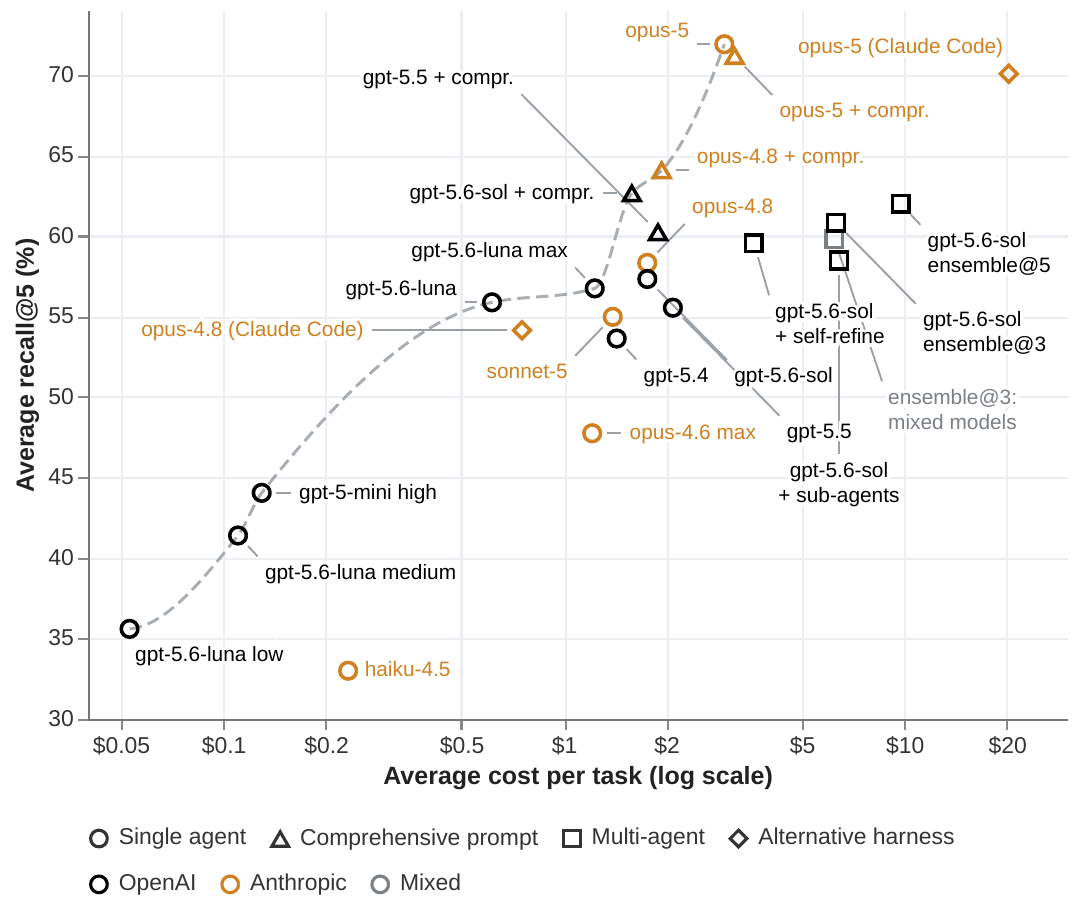}
        \caption{}
        \label{fig:pareto-front}
    \end{subfigure}
    \vspace{-12pt}
    \caption{Frontier models' review capabilities have steadily increased over six months in 2026, with \texttt{claude-opus-5} as the best-performing model (Left). The performance-cost Pareto front is dominated by small OpenAI models (\texttt{gpt-5.6-luna}) for cheaper choices, and large Anthropic models (\texttt{opus-4.8}, \texttt{opus-5}) for most-performant options. Multi-agent configurations provide improvements at a large cost (Right).}
    \label{fig:combined-results}
    \vspace{-10pt}
\end{figure}

We then use \bench{} to systematically study agent review quality. 
First, we evaluate standalone review agents with different models, and ask: \textbf{How well do review agents detect PR bugs (RQ1)?}
We found that review agents have a good success rate on detecting bugs in GitHub PRs: 
The best review agent (with \texttt{claude-opus-5-xhigh}) catches \BestCmtRecall\% of bugs that humans found before merge, and \BestFixRecall\% of bugs that humans found after merge, \textit{within top-5 review comments}.
This bug-catching review capability has steadily increased over model evolution, going from \EvolRecallOld\% (\texttt{opus-4.6}) to \EvolRecallNew\% (\texttt{opus-5}) from Feb to July 2026 (Figure~\ref{fig:performance-time-trend}). 
However, despite the promising trend, agents still miss 20\%+ bugs human found, indicating a need for human-in-the-loop review. 

We corroborate our results with \bench's proprietary portion built from industry repositories that are (1) industry-scaled and (2) certainly not part of models' training data.
We found that review agents struggle much more with the proprietary portion, dropping up to \AdoDrop\% performance, but with a similar promising trend of increased performance over model evolution (Figure~\ref{fig:performance-time-trend}).
The sharp drop is partially explained by different task distribution: 
tasks sourced from industry repositories contain far more bugs that are not directly inferrable from the diffs -- they require more project contexts to be confirmed.

Next, we study the conditions for successful agent reviews.
We collected features on repositories, PRs, and bugs, and ask: \textbf{How do different factors impact review quality (RQ2)?}
We modeled the probability that a review detected a bug using multivariable logistic regression, with our collected features. 
We found that review agents are less effective for catching bugs in larger and more heterogeneous PRs, or bug-fixing PRs.
Non-functional or less severe bugs are harder to detect, as well as bugs that require non-local reasoning.
This shows that review quality is heterogeneous, and developers need to form calibrated trust on review agents based on the PR context.

We then examine agent trajectories, and compare the errors different agents have made, and ask 
\textbf{Where do agents spend time and why do they fail (RQ3)?}
We found that agents spend most of their turns reading and searching code but rarely execute it, with the best model \texttt{claude-opus-5} executing code in \ExecuteLong\% of turns (vs. \texttt{gpt-5.6-sol} in \ExecuteShort\%).
Agents also differ substantially in their \textit{persistence}: 
half of \texttt{claude-opus-5} runs end by turn \HalfLifeLong{}, compared with turn \HalfLifeShort{} for \texttt{gpt-5.6-sol}.
Our qualitative analysis also reveals cases where agents trust misleading PR descriptions or code comments, or fail to account for repository-wide behavior and external library semantics, reflecting limitations of current frontier models.

Finally, we experiment with various agent design dimensions, and ask: \textbf{What agent design choices improve bug-detection performance (RQ4)?}
We found that scaling compute is effective, and most effective through \textit{prompt} changes.
Exploration-nudging prompts improve \PromptGain\% performance (\PromptCostRatio x cost), while multi-agents improve \MultiAgentGain\% performance (\MultiAgentCostRatio x cost).
Increased reasoning efforts can improve model performance from \EffortLowRecall\% to \EffortMaxRecall\% at \EffortCostRatio x cost.
There is a clear trade-off to make in the cost-performance frontier (Figure~\ref{fig:pareto-front}).

Based on empirical findings from our benchmark, we make recommendations for developers to adopt code review agents (Section~\ref{sec:discussion}).
We recommend that developers scale their (automated and human) review efforts based on PR complexity, and steer agent review directions based on project or PR priorities. 
They should also incorporate repository knowledge for better review quality, especially when they work on large proprietary repositories.
If cost is a concern, they should consider dispatching different models and agent configurations based on PR characteristics. 
With evolving model behaviors, it would be a good idea to monitor review quality and adapt the harness periodically.
We also recommend developers to apply the methodology of \bench{} and build their own \textit{customized} benchmark of their software repositories.

To summarize, this paper contributes: 
\begin{itemize}
    \item A scalable data construction and evaluation pipeline that transforms a repository into a suite of PR review tasks with human-verified bugs (Section~\ref{sec:data-collection}).
    \item \bench{}, a constructed benchmark from open-source repositories (Section~\ref{sec:data-collection}). 
    \item A comprehensive evaluation of review agents, with an analysis of when review agents work and when they do not (Section~\ref{sec:evaluation}).
    \item Actionable guidance for developers to adopt review agents (Section~\ref{sec:discussion}).
\end{itemize}

\section{Background and Related Work}

\subsection{Modern code review}
Code review originated in formal, synchronous inspection processes with
prescribed roles and meetings~\cite{fagan1976design}. 
\emph{Modern code review} is a more lightweight, asynchronous process, usually involving rapid feedback from a small number of reviewers~\cite{rigby2013convergent,bacchelli2013expectations, sadowski2018modern}. 
There is empirical evidence that code review contributes to software quality, but its outcomes are heterogeneous. 
Greater review participation is associated with fewer post-release defects~\cite{mcintosh2014impact}, while the usefulness of a particular comment depends on both its technical content and its context.
Large-scale studies show that useful feedback includes defects, design alternatives, and maintainability guidance, and that larger changes tend to receive a lower proportion of useful comments
~\cite{bosu2015characteristics,turzo2024useful}. 
We intend \bench{} to focus on the more objective aspects (bug-finding) of code review, but leave the more subjective or domain-specific aspects to human reviewers.

\subsection{Benchmarking code review}
The most popular approach to evaluate code review is to collect human review comments and evaluate generated comments against them, with text-overlap metrics such as BLEU or CodeBLEU or LLM-based evaluators:
Early benchmarks~\cite{tufano2021towards,tufano2022using, li2022automating} construct method-level code paired with human review comments, 
while more recent benchmarks like SWR-Bench~\cite{zeng2025benchmarking}, ContextCRBench~\cite{hu2026contextcrbench}, and c-CRAB~\cite{zhang2026code} have moved to full-project, PR-level evaluation.
In particular, c-CRAB converts human comments into executable tests and measures whether a separate coding agent can turn a review into a test-passing repair. 
These benchmarks are grounded primarily in reviewer comments before PR merge, and their reported scores can artificially penalize valid review findings that the original reviewers missed.
An alternative approach is to use LLMs to help augment and evaluate code review.
AACR-Bench~\cite{zhang2026aacr} augments reviewer comments with human-validated LLM-generated review comments, but this approach cannot assess performance on bugs that LLMs fail to identify in the first place.
DeepCRCEval~\cite{lu2025deepcrceval} and CRScore~\cite{naik2025crscore} use LLM-based evaluators to evaluate various aspects of code review (e.g., conciseness, comprehensiveness), but their evaluations are limited by the reliability of LLM evaluators and do not measure historical confirmed-bug recovery.

Our benchmark takes a complementary approach:
we identify bugs verified by human-authored change, including defects that were caught or escaped the original review,
and then evaluate directly whether a review agent can identify that defect in the introducing PR. 
This provides a recall-oriented test of whether review agents recover known, consequential issues across realistic repository contexts,
instead of a test of textual similarity or a general measure of comment style.
Our approach connects to the established literature on post-release defects and defect prediction~\cite{bird2011ownership,dambros2010comparison}, and similar to recent work that evaluates code review using discovered defects: 
\citet{lu2025practical} reconstruct introducing merge requests from industrial fault reports, while CR-Bench traces SWE-bench fixes to introducing PRs and evaluates reviews against the known bugs~\cite{pereira2026cr}. 
Compared to these works, we contribute a scalable recipe to curate diverse bugs and task environments from software repositories, and apply the recipe to create a benchmark grounded both in public and proprietary repositories.

\subsection{Automating code review}
Automation has long supported individual parts of code review, including static analysis, reviewer recommendation, change prioritization, and suggested revisions. 
For example, Review Bot combined static analysis with reviewer recommendation to surface findings inside the review workflow~\cite{balachandran2013reducing}. 
Neural approaches later learned edits or comments from historical review data~\cite{tufano2021towards,tufano2022using,li2022automating}. 
LLM-based systems extend this line of work to open-ended agentic review that can inspect repository context, use development tools, and produce several candidate findings~\cite{tang2024codeagent, tantithamthavorn2026rovodev}.

Recent evidence nevertheless cautions against over-trust of LLM-generated reviews. 
A controlled study with expert developers found that LLM-generated reviews redirected reviewers' attention and increased the number of low-severity issues found, but did not increase detection of high-severity issues, save review time, or improve confidence~\cite{tufano2025doubleedged}. 
An industrial study of 4,335 pull requests found that developers acted on many automated comments, while also reporting faulty, unnecessary, and irrelevant feedback and observing longer average PR closure times~\cite{cihan2025automated}.
A recent roadmap consequently frames the field around context-aware, value-driven, human-AI review rather than replacement of human reviewers~\cite{yang2026roadmap}. 
Our work advances this agenda by evaluating end-to-end review agents on bug detection, characterizing the change and repository conditions under which they fail, to inform when to allocate human oversight.

\section{Data Collection}
\label{sec:data-collection}

\begin{figure}
    \centering
    \includegraphics[width=1\linewidth]{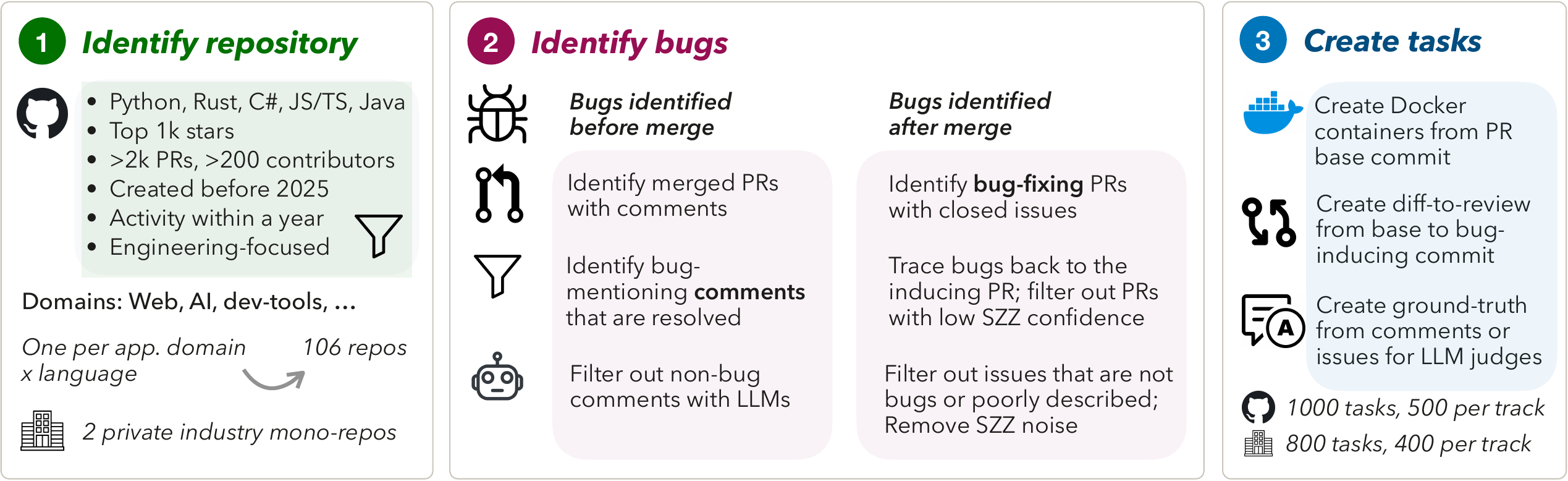}
    \caption{\bench{} tasks are created from human-verified bugs in repository history. We identify bugs from human comments and bug-fixing PRs, trace them back to the inducing PR, and create containerized tasks.}
    \label{fig:data-pipeline}
\end{figure}

In this section, we describe our data collection methodology, from how we select repositories, how we identify human-verified bugs, to how we create task environment for review agent evaluation (an overview is provided in Figure~\ref{fig:data-pipeline}).

\subsection{Repository selection}
We selected five popular programming languages: Python, Rust, C\#, JavaScript/TypeScript, and Java, based on StackOverflow's survey~\cite{stackoverflow2025technology}.
For each language, we identified the most popular, well-maintained engineering repositories with substantial activities.
We first identified top-1{,}000 starred GitHub repositories, and then filtered down to ones with more than 2{,}000 pull requests (PRs) and more than 200 contributors.
We further filtered out repositories that are not engineering-focused (e.g., tutorials), that are created after 2025, or have no activities within a year, to focus on established, actively maintained software projects.
From the remaining repositories, we annotated their application domains (e.g., AI/ML, databases, security, DevOps, web frameworks, gaming, embedded, etc.) with an LLM, manually verified that the annotations are reasonable, and then sampled one repository per domain for each language. 
This gives us a final of 106 repositories\footnote{We collected the repositories in June 26th, 2026.} spread across 31 application domains (statistics in Figure~\ref{fig:repo-stats}).

\subsection{Bug identification}
\paragraph{Bugs identified before merge.}
First, we identify bugs that developers have caught during PR review -- i.e., they are mentioned in PR discussion and fixed before merge. We collect these tasks in 2 steps:

\begin{enumerate}
    \item \emph{Mining.} We first collect all merged PRs and their discussion threads from GitHub API. We filter out PRs without any review threads.
    This yields \MinedPrs{} PRs with \MinedThreads{} review threads.
    We then apply deterministic quality filters to remove low-quality comments,
    filtering out comments that are bot-authored, self-authored, too short, on non-source-code files, or are rejected.
    After filtering, \RetainedThreads{} candidate threads remain across 117{,}522 PRs.

    \item \emph{Sampling and LLM filtering.} Because the surviving population is
    far larger than what we can evaluate, we draw a stratified sample of 90 threads per repository. 
    Within a repository we order PRs by merge time, split them into
    10 equal-count temporal strata, and make sample spans equally across the repository's history.
    This yields 8{,}639 threads.
    We then use an LLM to further label and filter comments that are not bugs (e.g., style, naming, tests, nits, refactors, and concerns the replies show were clarified or declined).
    This results in 865 vetted bugs across 99 repositories.

\end{enumerate}

\paragraph{Bugs identified after merge.}
Second, we identify bugs that developers caught after a PR is merged -- i.e., they are mentioned in a later issue and fixed in a later PR. We collect these tasks in 4 steps:

\begin{enumerate}
    \item \emph{Mining.} We first collect all merged PRs with fix-related keywords in their titles or labels.
    This yields \MinedFixPrs{} candidate PRs.
    We then identify closing issues for each PR, and remove the ones without substantive descriptions (less than 80 characters),
    yielding \MinedFixPrsWithIssue{} fix PRs.

    \item \emph{Sampling and LLM filtering.} 
    Because the surviving population is far larger than what we can evaluate, we again draw a stratified sample of 200 PRs per repository. 
    We then have an LLM label and filter issues that are not bugs and hence unpreventable (new features, refactors, documentation, performance work).
    This yields 10{,}784 fix PRs.

    \item \emph{Linking to bug-inducing PR.} We then trace each defect back to its inducing PR with SZZ~\cite{sliwerski2005changes}: 
    We identify the deleted/modified lines
    and use \texttt{git blame} to obtain the commits that last touched them; and map those commits to PR numbers using the repository's referencing convention.
    The primary inducing PR is the one accounting for the largest share of
    blamed lines. 
    This yields 8{,}846 fix--inducing pairs.
    We keep the top 12 pairs per repository ranked by SZZ blame share, resulting in 1{,}168 pairs.
    
    Since SZZ is noisy (a refactor that last-touched a line can be attributed as if it had introduced the defect), we further use an LLM (\texttt{claude-opus-4.8}) for quality control.
    The LLM is asked to classify the type of the inducing change and whether the blamed
    lines introduce the bugs, yielding 736 pairs.

    \item \emph{Ground truth screening.} Finally, we filter out poorly-described issues with an LLM. 
    We drop issues with no reported faulty behaviors (screenshot-only reports, unfilled templates, bare one-liners), non-bug issues that are vague or off-target, and issue/fix mismatches. 
    This yields a final set of 689 fix--inducing PR pairs.

\end{enumerate}

\textit{Data quality control.}
We iterated our bug identification pipeline over multiple weeks, through validating the quality of curated bugs and adding additional deterministic or LLM filters. 
We randomly sampled 40 identified bugs from the final pipeline and assessed whether each reported bug
(i) represented a concrete defect and
(ii) could be detected by reviewers using the available context.
A task was labeled valid only if both criteria were satisfied.
Overall, 37 tasks (92.5\%) were valid, with an approximate 95\% Wilson
confidence interval of 80.1-97.4\%.

\subsection{Task creation}
From the identified and verified bugs and bug-inducing PRs, we construct our review tasks.
For each PR, we identify the \emph{base commit} and the \emph{reviewed commit}.
For bugs identified before merge, the reviewed commit is the commit that carries the ground-truth bug comments.
For bugs identified after merge, the reviewed commit is the final commit before merge.
We then construct the base-to-reviewed-commit diff for review.
We create a DockerFile per task and remove ones with build failures. 
We then sample the tasks down to 500 per track.

\textit{Task format.} Each task instance contains a Docker snapshot of the repository at the inducing PR's base commit, with history through that commit intact but all later references deleted to prevent leakage. 
The PR title and description are provided as context, and the change is provided both as a file and uncommitted working-tree modifications.
The agent reviews the change without being told that a bug exists and writes its review to \texttt{review.md}.
The review is structured as a list of review items, self-ranked by the agent based on severity, following popular review agent design~\cite{anthropicCodeReview}.
We restrict Internet access and explicitly prompt the agent to avoid cheating behaviors.
We later conducted an AI-assisted auditing of agent trajectories and did not find any cheating behaviors.

\textit{Task scoring.}
Each review finding is scored by an LLM-as-judge panel against the ground truth (bug-mentioning comments or issues) and annotated if there is a match.
To mitigate judge biases towards same-family models and single-judge instability~\cite{zheng2023judging},
we use three judge models: \texttt{gpt-5.5}, \texttt{claude-opus-4.8}, and \texttt{gemini-3.1-flash}, and take the majority vote as the final result.
We independently annotated 40 samples of agent reviews and found that we achieved almost perfect human-LLM agreement ($\kappa=0.89$). 
We also observed that judges are fairly consistent with each other, achieving almost perfect across-LLM agreement (Fleiss' $\kappa=0.95$).

For evaluation, we measure recall@k as the proportion of tasks for which the ground-truth issue is identified among the agent's top-k findings. Varying k captures different limits on developer attention and team review budgets.

\subsection{Benchmark characteristics}

\begin{figure}
    \centering
    \includegraphics[width=\linewidth]{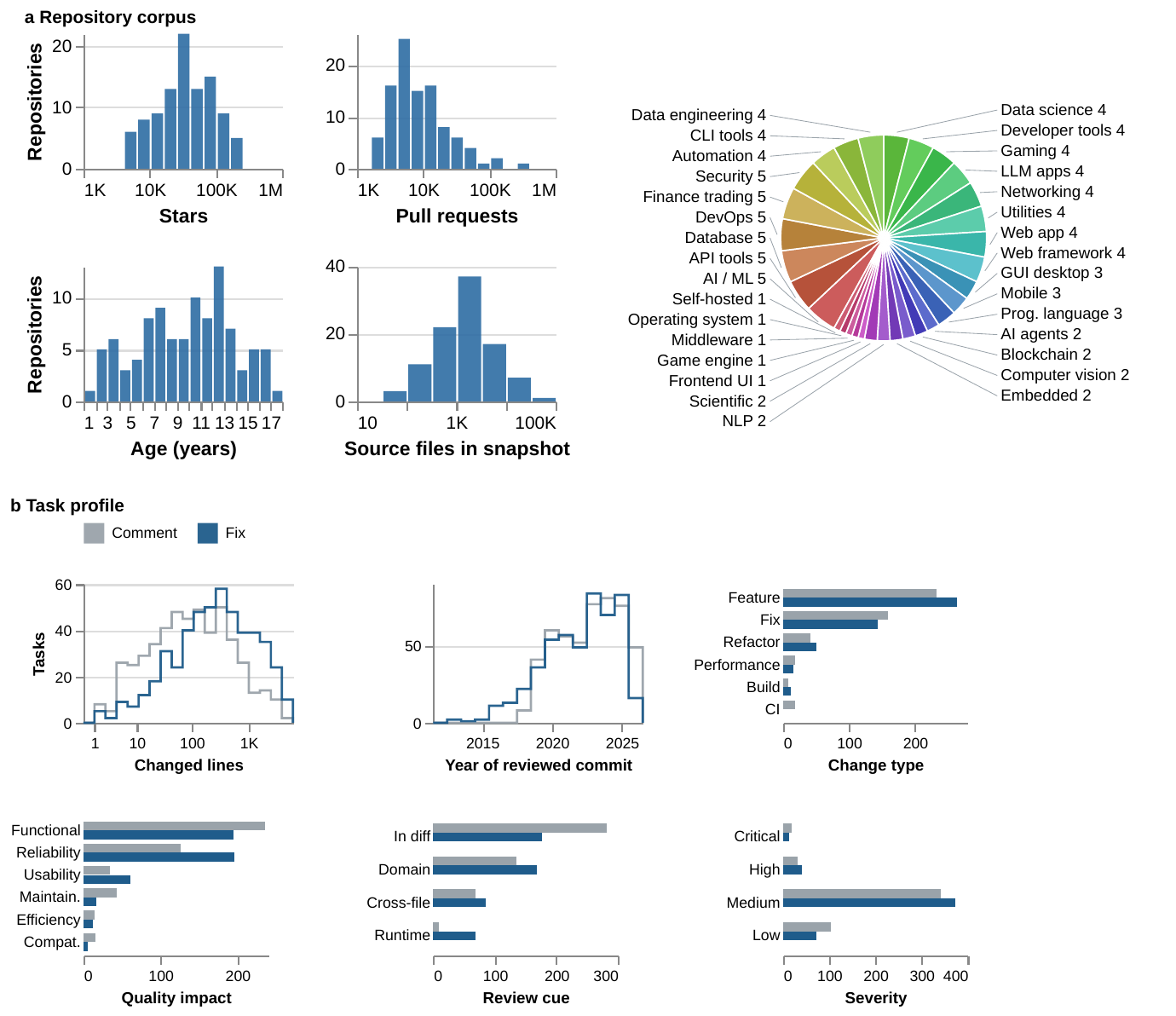}
    \caption{Statistics of collected repositories.}
    \label{fig:repo-stats}
\end{figure}

Applying our data pipeline to GitHub repositories, we collect \bench, with \BenchTasks{} tasks across \BenchRepos{} popular and established open-sourced repositories (Figure~\ref{fig:repo-stats}).
The tasks are spread out from 2015 to 2026, with most reviewed changes single-language (\SingleLanguageShare\%) and a median churn of \MedianChurn{} lines. 

\textit{Task annotations.}
We in addition annotate various task attributes for downstream analysis.

\begin{enumerate}
    \item \emph{Code metrics.} We collect diff size, file kinds, languages, and repository size.
    We also capture file-level dependency graph with CodeQL and compute the number of dependent / dependency files.

    \item \emph{LLM annotations.} We further create rubrics and use an LLM to annotate various semantic attributes of code change and bugs:
    We annotate \emph{change type} (following taxonomy in~\citet{li2025rise}), \emph{bug impact} based on ISO/IEC~25010 quality impact~\cite{iso25010_2011}, 
    plus \emph{bug severity} and \textit{bug visibility}. 
    We validated a sample of 40 annotations and found that we have almost perfect human-LLM agreement for change type and bug impact ($\kappa=0.88, 0.82$ respectively), and substantial agreement for bug severity and visibility ($\kappa=0.64, 0.68$).
    For severity, we group critical / high and medium / low in our analysis, which yields a higher $\kappa=0.90$.

\end{enumerate}

\subsection{Proprietary benchmark construction}
Besides the public benchmark, we applied the same data pipeline on two industry mono repositories on enterprise platforms for online collaboration and productivity, collecting 800 additional review tasks.
These tasks are sourced from repositories with an average of >100k source files, which is a much larger scale compared to open-source repositories (an average of 3k source files).

\section{Experiments}
\label{sec:evaluation}
Next, we use our benchmark to evaluate review agents' bug-detection performance.

We first evaluate across different model configurations to understand \textit{how well review agents detect
PR bugs} (RQ1).
We then use the collected data to do a large-scale empirical analysis, to understand \textit{what factors impact review agents' bug detection capabilities} (RQ2).
We then analyze the agent trajectories to understand \textit{where do agents spend time and why do they fail} (RQ3).
We finally evaluate with different harness variations, to understand \textit{how agent design choices impact bug detection} (RQ4).

\subsection{Research Design}

\paragraph{Evaluating and comparing agentic models (RQ1).}
We first evaluate across a wide range of large and small models with strong coding capabilities:
\begin{itemize}[noitemsep,topsep=0pt, leftmargin=2em]
    \item \textbf{Large frontier models}: \texttt{claude-opus-5}, \texttt{claude-opus-4.8}, \texttt{claude-opus-4.6}, \texttt{gpt-5.6\-sol}, \texttt{gpt-5.5}, and \texttt{gpt-5.4}. These represent large, strong coding models available at the time of our experiments.
    \item \textbf{Smaller and older models}: \texttt{claude-sonnet-5}, \texttt{gpt-5.6-luna}, \texttt{claude-haiku-4.5},
    and \texttt{gpt-5\-mini}. These represent decent coding models that are cheaper and smaller.    
\end{itemize}

To control for variations in agent harnesses, we use \texttt{copilot-cli}, pinned to 1.0.74, as the agent harness to evaluate and compare different models.\footnote{We ablate this choice with \texttt{claude-code} in RQ4 and observed better performance from \texttt{copilot-cli}.}
For the review prompt, we use the official prompt provided by \texttt{copilot-cli} in the \texttt{/review} command.
For all models, we use a reasoning effort of \texttt{xhigh} when available.
Otherwise, we use the max available reasoning effort for the model.

In addition to the main performance metric recall@k, we also report (1) cost metrics, including token usage, estimated cost (in USD), and latency, and (2) review length metrics, including number of findings and words.

For the proprietary benchmark portion, we evaluate four top-performing models: \texttt{claude-opus-5}, \texttt{claude\-opus-4.8},  \texttt{gpt-5.6-sol}, and \texttt{gpt-5.5}.
We compare their results to the open-source portion of our benchmark.

\paragraph{Analyzing variation in agent performance (RQ2).}
To identify the conditions under which review agents are more or less successful, we model the probability that a review detects a bug using multivariable logistic regression. 
Because the literature on LLM-based code review is still nascent, we draw on the well-established literature on human code review to motivate our hypotheses, under the assumption that factors known to make changes harder for human reviewers are likely to affect review agents in similar ways.
Specifically, we draw on findings that larger changes are reviewed less effectively and attract fewer useful comments~\cite{bosu2015characteristics, baum2019associating, ram2018reviewability}, and composite changes that mix multiple concerns are harder to review~\cite{tao2012understand, barnett2015decomposition}.
Review quality also varies with the kind of artifact under review: 
test code receives less review discussion than production code~\cite{spadini2018testing}, and build and configuration files attract fewer useful comments~\cite{bosu2015characteristics}.
Finally, studies show that \emph{delocalized} defects that require reasoning about code outside the inspected artifact, are substantially harder to find~\cite{dunsmore2000delocalisation, baum2019associating}, and the majority of issues raised in code review concern maintainability rather than functional defects~\cite{mantyla2009types, beller2014problems}.
Motivated by these findings, we test the following hypotheses:

\begin{itemize}[noitemsep,topsep=0pt, leftmargin=3em]
    \item[\textbf{H1:}] \textbf{Detection probability decreases with higher PR change complexity.}
    We measure change complexity with number of change lines, the number of dependent files, and the number of dependency files.

    \item[\textbf{H2:}] \textbf{Detection probability is associated with PR change type.}

    \item[\textbf{H3:}] \textbf{Detection probability is associated with PR change language.}

    \item[\textbf{H4:}] \textbf{Detection probability decreases with higher PR change heterogeneity.}
    We measure change heterogeneity with the number of change types within one change (e.g., on source code, tests, build files, etc.), and with the file extension entropy.

    \item[\textbf{H5:}] \textbf{Detection probability decreases with higher repository complexity.}
    We measure repository complexity with the number of lines in the repository snapshot.

    \item[\textbf{H6:}] \textbf{Detection probability increases for diff-visible bugs.}

    \item[\textbf{H7:}] \textbf{Detection probability is associated with bug impact.}

    \item[\textbf{H8:}] \textbf{Detection probability is associated with bug severity.}

\end{itemize}

All predictors are entered simultaneously, with continuous measures standardized where applicable. 
We also include the model used for review, task track, and calendar time as adjustment covariates. 

\vspace{-10pt}
\[
\scriptsize
\begin{aligned}
\texttt{review\_caught} \sim{}
&\texttt{log\_num\_changed\_lines\_z\_score}
+ \texttt{log\_dependent\_files\_z\_score} 
+ \texttt{log\_dependency\_files\_z\_score}\\
&+ \texttt{log\_repository\_nloc\_z\_score}
+ \texttt{num\_subchange\_types}
+ \texttt{extension\_entropy\_z\_score} \\
&+ \texttt{bug\_is\_local\_visible} 
+ \texttt{high\_severity\_bug}
+ C\!\left(
  \texttt{change\_type},
  \operatorname{Treatment}(\texttt{ref}=\texttt{feature})
\right) \\
&+ C\!\left(
  \texttt{language},
  \operatorname{Treatment}(\texttt{ref}=\texttt{Python})
\right)
+ C\!\left(
  \texttt{bug\_iso\_impact},
  \operatorname{Treatment}(\texttt{ref}=\texttt{Functional Suitability})
\right)\\ 
&+ C(\texttt{model\_key})
+ C(\texttt{track})
+ C(\texttt{calendar\_time}) 
\end{aligned}
\]

Because a task could have multiple review executions, observations from the same task may be correlated. We therefore cluster standard errors by task and applied finite-sample and degrees-of-freedom corrections.
For each scalar predictor, we report the adjusted odds ratio, (\(\exp\{\hat{\beta}\}\)), and its 95\% cluster-robust confidence interval. 
The adjusted odds ratio describes the change in the odds of the outcome associated with a one-unit increase in the predictor, holding the other predictors constant. 
We test each scalar predictor using a one-degree-of-freedom cluster-robust Wald (F)-test. 
For each categorical predictor, we use an omnibus cluster-robust Wald (F)-test to assess whether its levels are jointly associated with the outcome. 
We report adjusted odds ratios and 95\% confidence intervals for individual levels relative to the reference category.

We hypothesize that differences among models vary across data subsets. 
To test this hypothesis, we fit an additional regression model with interactions between model and the other factors. 
We further examine whether the associations between these factors and bug detection differ between human reviewers and LLMs. 
For this analysis, we fit a model incorporating human-detection observations identified from bug sources and test interactions between \texttt{is\_human\_review} and the other factors.

To account for multiple testing, we apply the Benjamini-Hochberg procedure~\cite{benjamini1995controlling} to control the false discovery rate at 5\%. 
We also confirmed that the fitted model converged, its design matrix was full rank, and that multicollinearity was within our specified thresholds (condition number < 30, the maximum variance inflation factor < 10).

\paragraph{Analyzing agent trajectories (RQ3).}
To characterize \textit{how} agents spend their budget, we instrumented the trajectories of all executions.
Each tool call is classified into one of six actions by tool name, and, for shell calls, by parsing the command into segments and mapping the head verb:
\textit{Read} (open a file), \textit{Probe} (search or list),
\textit{Write} (create or edit), \textit{Execute} (run tests, builds, or ad-hoc scripts), \textit{Git} (inspect history or diffs), and \textit{Other}.

To characterize \textit{why} agents fail, we analyze the six frontier models and decompose failures into three groups,
by matching file paths explored in the trajectory and the ground truth: 
whether the agent inspected the file holding the ground-truth defect, 
whether the review discussed that file, and whether the discussion was judged a correct diagnosis.
We then complement this with an inductive open coding of 20 sampled tasks with diverging review agent outcomes.
Each case was inspected and annotated on how the agents succeed or make mistakes.

\paragraph{Evaluating and comparing agent harnesses (RQ4).}
We finally evaluate across various agent configurations, testing how the choice of harness, prompt, and orchestration impacts agent performance.

\begin{itemize}[noitemsep,topsep=0pt, leftmargin=2em]
    \item \textbf{Reasoning efforts.} We ablate reasoning efforts for \texttt{gpt-5.6-luna}, with reasoning efforts set to low, medium, and max.

    \item \textbf{Harness.} We ablate the harness with \texttt{claude-code} for \texttt{claude-opus-4.8} and \texttt{claude-opus-5} using its \texttt{/review} command.
    
    \item \textbf{Prompt.} We ablate the default \texttt{copilot-cli} review prompt with a prompt that encourages comprehensiveness. The default prompt optimizes for signal-to-noise ratios (i.e., models are encouraged to only output important findings).
    
    \item \textbf{Orchestration.} We ablate the default single-agent loop with three ways to scale up compute efforts.
    The first is through \textit{self-refine}, where we have a second agent to improve the reviews from the first pass.
    The second is through \textit{candidate-ensemble}, where we have an orchestrator agent that collects several independent candidate reviews and consolidates them into one set of findings. We vary both how many candidates are pooled (three or five) and whether those candidates come from the same model or from different models.
    In the case of different models, we sample candidates from \texttt{claude-opus-4.8}, \texttt{gpt-5.5} and \texttt{gpt-5.6-sol}. 
    The last is through \textit{sub-agents}, where we have an orchestrator agent that creates sub-directions to explore, and invoke sub-agents on demand, and create consolidated findings.
\end{itemize}

\subsection{Threats to Validity}

\textit{Construct validity.}
Our benchmark only measures review agents' bug detection capabilities.
Our results should therefore be interpreted as recovery of known bugs rather than as overall review usefulness.
Our evaluation uses LLM judges, which may be unreliable~\cite{zheng2023judging}.
We mitigate this by using a three-model panel and calibrate human-LLM agreement.

\textit{Internal validity.}
Despite multi-stage filtering, our benchmark construction can still yield incorrect ground truths or unsolvable tasks.
We mitigate this with extensive data quality control and manual verification of samples. 
Agent behavior is also stochastic and sensitive to changes, which we mitigate with pinning the primary harness and holding the prompt and task environment fixed.
Our observational regressions estimate adjusted associations, not causal effects.
Unmeasured factors and LLM annotation errors may confound estimates or attenuate effects.

\textit{External validity.}
We cover popular open-source repositories across domains and two large industry monorepos.
These choices improve diversity but do not represent all software development.
Our selection favors mature, active, popular projects and source-code changes with detailed textual evidence and reproducible environments.
Results may differ for smaller or younger projects, uncommon languages, generated or configuration-heavy code, and other review practices.

\begin{table*}[t]
\centering
\scriptsize
\renewcommand{\arraystretch}{1.08}
\setlength{\tabcolsep}{2.5pt}
\caption{Bug-detection performance and cost across model configurations on GitHub repositories.}
\label{tab:majority-recall-at-k}
\begin{tabular}{llrrrrrrrrrrr}
\toprule
\multirow[t]{3}{*}{\textbf{Run}} & \multirow[t]{3}{*}{\textbf{Track}} & \multicolumn{4}{l}{\textbf{Recall}} & \multicolumn{2}{l}{\textbf{Review}} & \multicolumn{3}{l}{\textbf{Tokens}} & \textbf{Cost} & \textbf{Latency} \\
\cmidrule(lr){3-6}\cmidrule(lr){7-8}\cmidrule(lr){9-11}
 &  & \textbf{@1} & \textbf{@3} & \textbf{@5} & \textbf{@10} & \textbf{Count} & \textbf{Words} & \textbf{In} & \textbf{Cache} & \textbf{Out} & \textbf{(\$)} & \textbf{(min)} \\
 &  &  &  &  &  &  & \textbf{(K)} & \textbf{(M)} & \textbf{(M)} & \textbf{(K)} &  &  \\
\midrule
\multirow[t]{2}{*}{\texttt{claude-opus-5}}
  & Cmt. & 48.3 & 71.2 & 79.6 & 86.0 & 5.76 & 1.08 & 2.43 & 2.34 & 34.3 & 2.55 & 8.6 \\
  & Fix & 36.2 & 55.4 & 64.2 & 68.6 & 6.78 & 1.28 & 3.37 & 3.27 & 43.9 & 3.38 & 11.1 \\
\addlinespace[0.1em]
\multirow[t]{2}{*}{\texttt{claude-opus-4.8}}
  & Cmt. & 45.6 & 65.9 & 70.0 & 71.5 & 2.71 & 0.61 & 1.23 & 1.17 & 24.0 & 1.54 & 6.5 \\
  & Fix & 32.0 & 44.6 & 46.6 & 47.8 & 2.84 & 0.67 & 1.66 & 1.59 & 30.4 & 1.98 & 7.9 \\
\addlinespace[0.1em]
\multirow[t]{2}{*}{\texttt{claude-opus-4.6 (max)}}
  & Cmt. & 45.4 & 58.7 & 59.1 & 59.3 & 2.08 & 0.36 & 0.90 & 0.86 & 12.5 & 0.99 & 4.1 \\
  & Fix & 27.0 & 35.0 & 36.4 & 37.0 & 2.46 & 0.45 & 1.40 & 1.35 & 17.5 & 1.44 & 6.0 \\
\addlinespace[0.1em]
\multirow[t]{2}{*}{\texttt{gpt-5.6-sol}}
  & Cmt. & 43.5 & 61.9 & 65.6 & 66.6 & 2.72 & 0.34 & 1.01 & 0.86 & 10.4 & 1.50 & 3.7 \\
  & Fix & 29.6 & 45.6 & 49.0 & 51.8 & 3.57 & 0.43 & 1.64 & 1.48 & 16.3 & 2.03 & 5.2 \\
\addlinespace[0.1em]
\multirow[t]{2}{*}{\texttt{gpt-5.5}}
  & Cmt. & 43.7 & 61.6 & 63.3 & 63.5 & 2.18 & 0.28 & 1.65 & 1.49 & 11.2 & 1.86 & 4.0 \\
  & Fix & 29.0 & 45.4 & 47.8 & 48.6 & 2.66 & 0.33 & 2.17 & 2.00 & 15.3 & 2.33 & 5.3 \\
\addlinespace[0.1em]
\multirow[t]{2}{*}{\texttt{gpt-5.4}}
  & Cmt. & 42.6 & 59.5 & 61.8 & 62.0 & 2.28 & 0.30 & 1.91 & 1.70 & 21.5 & 1.27 & 4.3 \\
  & Fix & 28.4 & 42.6 & 45.4 & 45.6 & 2.72 & 0.35 & 2.49 & 2.23 & 26.6 & 1.59 & 5.3 \\
\addlinespace[0.2em]
\midrule
\multirow[t]{2}{*}{\texttt{claude-sonnet-5}}
  & Cmt. & 47.8 & 62.5 & 64.7 & 65.7 & 2.65 & 0.69 & 3.26 & 3.18 & 35.1 & 1.18 & 9.1 \\
  & Fix & 30.6 & 43.2 & 45.2 & 45.6 & 3.04 & 0.80 & 4.56 & 4.46 & 45.9 & 1.60 & 11.7 \\
\addlinespace[0.1em]
\multirow[t]{2}{*}{\texttt{gpt-5.6-luna}}
  & Cmt. & 43.5 & 60.4 & 63.9 & 65.4 & 2.76 & 0.32 & 2.24 & 2.04 & 17.5 & 0.51 & 3.1 \\
  & Fix & 30.0 & 44.6 & 47.8 & 50.2 & 3.52 & 0.40 & 3.33 & 3.07 & 24.9 & 0.72 & 4.3 \\
\addlinespace[0.1em]
\multirow[t]{2}{*}{\texttt{claude-haiku-4.5 (none)}}
  & Cmt. & 33.0 & 41.2 & 43.0 & 43.8 & 2.58 & 0.44 & 0.91 & 0.87 & 13.4 & 0.21 & 2.2 \\
  & Fix & 17.6 & 21.8 & 23.0 & 23.2 & 3.16 & 0.52 & 1.20 & 1.15 & 16.0 & 0.26 & 2.7 \\
\addlinespace[0.1em]
\multirow[t]{2}{*}{\texttt{gpt-5-mini (high)}}
  & Cmt. & 38.4 & 52.7 & 54.7 & 55.3 & 2.72 & 0.54 & 1.40 & 1.14 & 13.5 & 0.12 & 5.3 \\
  & Fix & 21.4 & 31.2 & 33.4 & 33.8 & 3.48 & 0.71 & 1.57 & 1.30 & 19.4 & 0.14 & 5.4 \\
\bottomrule
\end{tabular}
\end{table*}

\subsection{How well do review agents detect PR bugs (RQ1)?}

First, we found that review agents are decently strong at detecting bugs (Table~\ref{tab:majority-recall-at-k}): 
The top-5 findings from \BestModel{} cover \BestCmtRecall\% bugs that are identified by humans before merge, and \BestFixRecall\% bugs that are identified after merge.
This only costs \$\BestCost{} per task, with an average latency of \BestLatency{} minutes.
We also observe a steady increase of review capability from frontier models: 
From February to July, the review agents have gone from catching \EvolRecallOld\% to \EvolRecallNew\% (with \texttt{opus-4.6} vs. \texttt{opus-5}, ***) of bugs, averaged across both pre-merge and post-merge bugs.

\begin{figure}[t]
    \centering
    \begin{subfigure}[t]{0.32\linewidth}
        \centering
        \includegraphics[width=\linewidth]{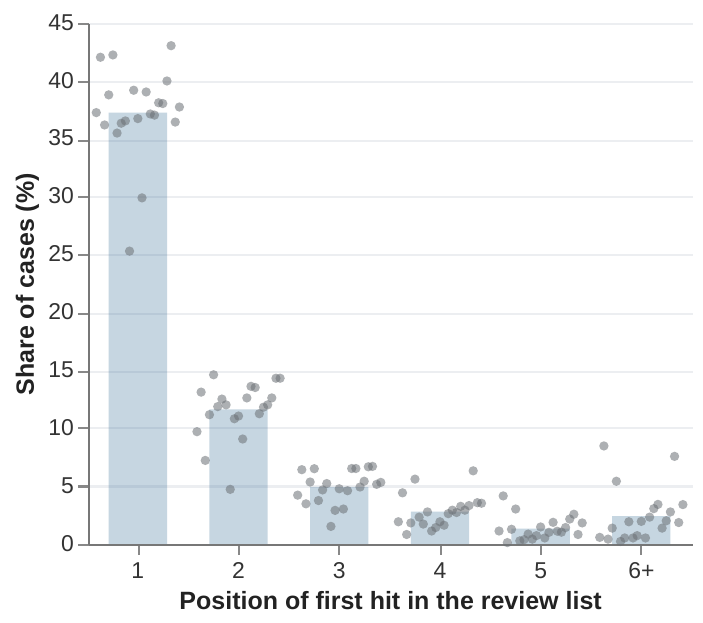}
        \phantomsubcaption
        \label{fig:hit-distribution}
    \end{subfigure}
    \hfill
    \begin{subfigure}[t]{0.28\linewidth}
        \centering
        \includegraphics[width=\linewidth]{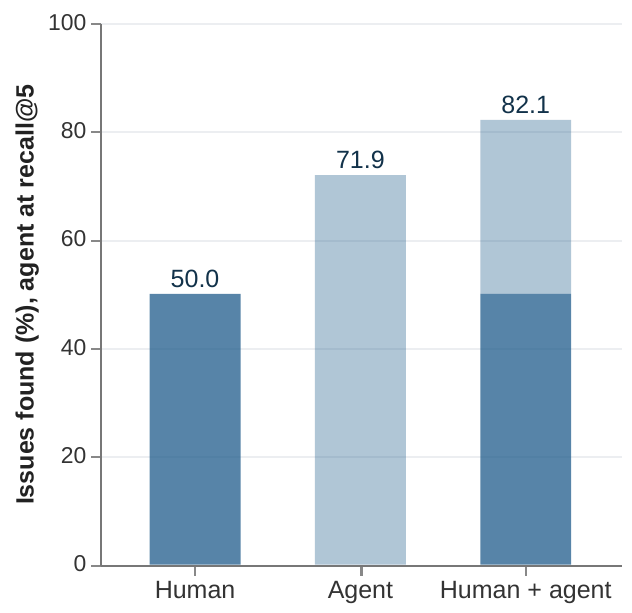}
        \phantomsubcaption
        \label{fig:human-agent}
    \end{subfigure}
    \hfill
    \begin{subfigure}[t]{0.36\linewidth}
        \centering
        \includegraphics[width=\linewidth]{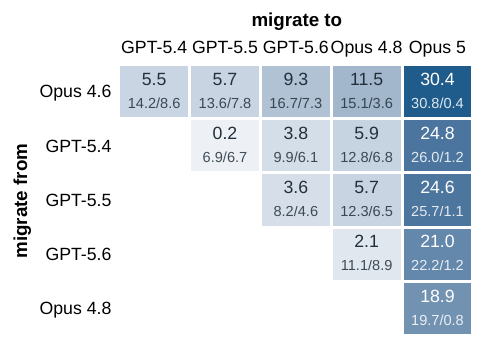}
        \phantomsubcaption
        \label{fig:migration-heatmap}
    \end{subfigure}
    \vspace{-15pt}
    \caption{Bug detection performance and model migration gains.
(a) Distribution of the first correct finding's position in the review list, with correct findings concentrated near the top.
(b) Detection rates for human review, agent review, and their combination, highlighting their complementary coverage.
(c) Pairwise gains from migrating between models, with darker shading indicating larger gains.}
    \vspace{-10pt}
    \label{fig:hit-and-coverage}
\end{figure}

Noticeably, the models are also fairly strong at ranking their findings: 
the findings that are a hit are heavily concentrated at the left end (\HitTopThree\% of findings ever found is found in top-3 items, Figure~\ref{fig:hit-distribution}).
This motivates us to design an improvement strategy, where we explicitly ask models to provide \textit{more} comprehensive findings and then \textit{rank} them by severity.
This reduces the risk of discarding valid bugs early, and leverages models' strong ranking capabilities for prioritizing the most consequential findings, which proves to be the strongest improvement (cf. Section~\ref{sec:rq4-agent-design}).

If we look across smaller and bigger models, we can also see a clear trade-off between performance and cost (Figure~\ref{fig:pareto-front}).
Some small models are very effective at finding bugs at their cost tier -- for example, \SmallModel{} matched \SmallRecallShare\% \BestModel{} performance at \SmallCostShare\% cost.
This suggests room to make deliberate model choices based on cost constraints.

\paragraph{Opportunities for human-agent collaboration.}
Our results indicate a clear opportunity for human-agent collaborative code review.
By the nature of our benchmark construction, humans alone only catch \HumanRecall\% of the bugs, while best agent alone catches \AgentRecall\%.\footnote{From our benchmark construction telemetry, we observe similar amount of pre-merge and post-merge bugs (49.7\% vs. 50.3\%). This distribution can change for repositories with more (or less) thorough code review policies.}
If they were to collaborate together and combine their findings, this can catch \UnionRecall\% of the bugs.
Humans and agents are also observed to complement each other on different PR subgroups, which we will discuss in next subsection.

\paragraph{Agents fall short at large industry repositories.}
Overall, we have seen a significant (up to \AdoDrop\%) performance decrease when we apply the same agent on our internal repositories (Figure~\ref{fig:performance-time-trend}).
As we will show later, this can be partially attributed to their different bug distribution:
Tasks in our internal repositories contain fewer diff-visible bugs, but require deeper domain knowledge, inspecting larger contexts, or gathering runtime evidence for successful diagnosis.

\mybox{
\faArrowCircleRight~\textbf{Key Findings (RQ1)}:
Latest agentic models like \BestModel{} are strong code reviewers, catching \BestRecall\% bugs within top-5 findings at \$\BestCost. 
Models' review capabilities have steadily increased over time, but gaps remain that call for human engagement. 
Agents struggle more with large industry repositories.
}

\subsection{How are task conditions correlated with review agent performance (RQ2)?}
\label{sec:rq2-conditions}

\begin{figure}[t]
    \centering
    \includegraphics[width=\linewidth]{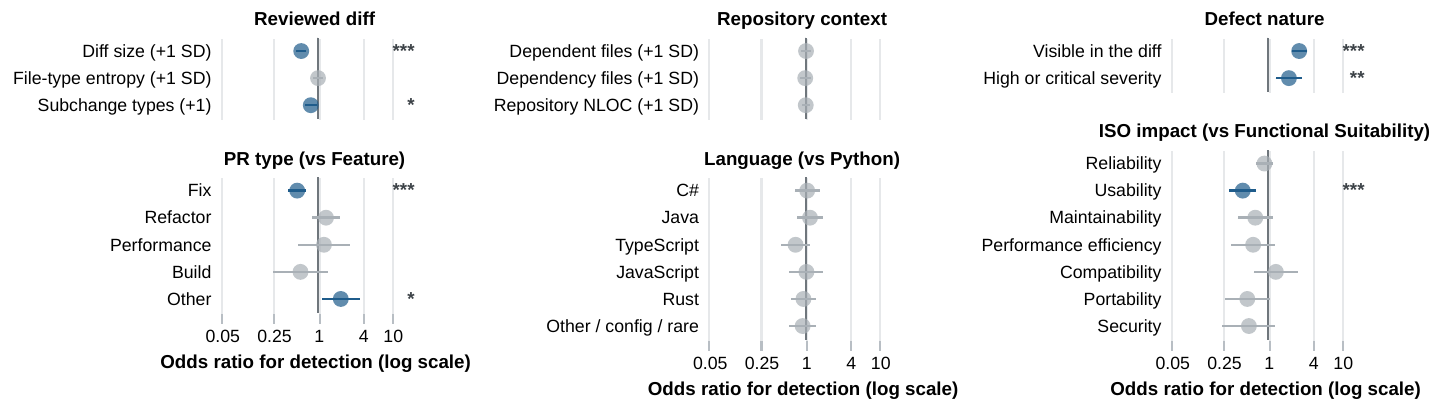}
    \caption{Odds ratios (log scale) with 95\% confidence intervals from the
    multivariable logistic regression of bug detection. Blue marks a
    significant effect: *** $p<0.001$, ** $p<0.01$, * $p<0.05$; grey is not
    significant.}
    \label{fig:condition_effect}
\end{figure}

\begin{figure}[t]
    \centering
    \includegraphics[width=1\linewidth]{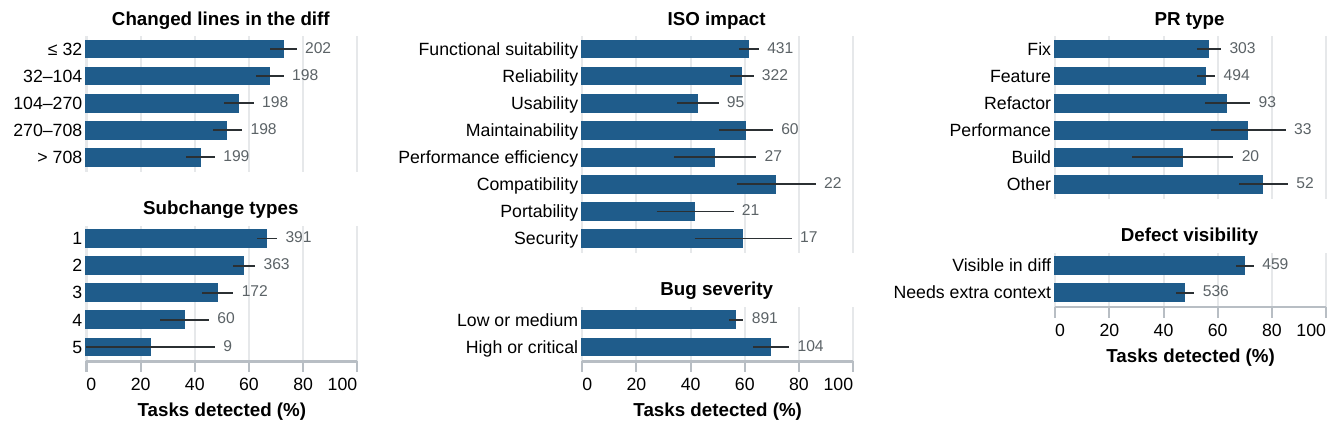}
    \caption{Bug detection rates by diff size, number of subchange types, ISO impact, bug severity, PR type, and defect visibility. Numbers beside bars indicate task counts.}
    \label{fig:condition_rates}
\end{figure}

\textit{Agents struggle with PRs with larger, more complex diffs (H1, H4).}
First, we observe that larger diffs significantly decrease detection probability (OR=\OrDiffSize, \SigDiffSize). 
This is expected, as larger diffs contain more code to review and hence dilute review attention, making bugs harder to detect.

Besides the volume, we also observe that more heterogeneous diffs are harder to review (OR=\OrHeterogeneity, \SigHeterogeneity). 
That is, when the diffs contain multiple types of changes (e.g., on source code, tests, configs), they are harder to review.
We hypothesize that this may come from that more complex changes require reviews to reason across modules and interfaces, hence harder to review.

\textit{Agents struggle with bug-fixing PRs (H2).}
We found that bug-fixing PRs (OR=\OrBugfixPR, \SigBugfixPR) are harder to review than other types of PR, when we adjust for other attributes (e.g., diff size) in regression.
This may be a result of when agents review a bug-fixing PR, their attention is diverted to verify the fix is plausible, with less attention to other regressions introduced.

Meanwhile, we do not observe any significant effects of PR language (H3). 
This may be because all the languages we study are popular enough that frontier models excel at them at a similar level.

Repository size (H5) also has no observed significant effect, when adjusted for diff size.
We hypothesize this is because, reviewing a given diff usually does not require extensive inspection and understanding of the whole repository contexts, but only the local contexts -- and the size of local contexts are correlated with diff size.

\textit{Agents struggle with less severe, non-functional bugs (H7, H8).}
Next, we look into how detection performance diffs across bug attributes.
We found that agents have harder time detecting non-functional bugs like usability (OR=\OrUsability, \SigUsability) bugs, with the same direction but weaker evidence for portability (OR=\OrPortability, $p$=\PPortability) bugs, but easier time with more severe bugs (OR=\OrSeverity, \SigSeverity).
This likely comes from how the agents are both trained and instructed to focus on high-impact issues, while less-critical issues can get neglected.

\textit{Agents struggle with bugs that are not directly visible (H6).}
We also found that agents succeed more with bugs that are directly visible in the diff (OR=\OrVisible, \SigVisible), compared to bugs that need additional project knowledge, cross-files check beyond the diff, or run-time checks.

\looseness=-1
Comparing GitHub tasks and industry tasks, we found that PRs in our internal repositories contain far fewer locally visible bugs (\AdoVisibleShare\% vs. \GithubVisibleShare\% in GitHub tasks), but contain many more bugs that require project context understanding.
This is a significant factor in explaining why models fall short more in industry tasks: adjusting for bug visibility alone explains \AdoLocalityExplained\% of the performance gap.
This may also reflect that industry repositories often require deeper domain knowledge that is not part of models' training data, and the models have to compensate through additional test-time search.

\paragraph{Model improvements are not uniform across PR types.}
We observe that model improvements are generally mixed across model pairs -- some net gains come at the cost of \MaxRegression\% of regressions (Figure~\ref{fig:migration-heatmap}).
One noticeable exception comes from \texttt{claude-opus-5}, which provides consistent improvements with few regressions (up to \BestRegression\%).
We also observe a significant interaction between models and PR types: \texttt{opus-5}'s edge over the other models is concentrated on feature and bug-fixing PRs, and narrows significantly on refactors (RoR=\OrTopOneRefactor, \SigTopOneRefactor) and other PR types (RoR=\OrTopOneOther, \SigTopOneOther).\footnote{The ratio of odds ratios (RoR) measures how models’ relative detection odds vary across conditions.}

\paragraph{Human reviews catch more maintainability issues.}
Comparing human reviews and agent reviews, we observe a significant interaction between bug impact and reviewer identity:
humans are much better (OR=\OrHumanMaintainability, \SigHumanMaintainability) at catching \textit{maintainability} issues than agent reviewers.
Yet overall, humans are also worse at identifying more severe bugs (OR=\OrHumanSeverity, \SigHumanSeverity) and reliability issues (OR=\OrHumanReliability, \SigHumanReliability).
This suggests that human-agent collaboration is also non-uniform: humans are better at keeping repositories more well-maintained, but agents can more tirelessly scrutinize for reliability issues and subtle yet critical bugs.

\mybox{
\faArrowCircleRight~\textbf{Key Findings (RQ2)}:
Agents struggle with larger, more heterogeneous, and bug-fixing PRs, and with non-functional, less severe bugs or those requiring non-local reasoning.
Differences in bug visibility explain \AdoLocalityExplained\% of the GitHub-industry performance gap.
We also observe complementary strengths for humans and agents: maintainability for humans, and severity and reliability for agents.
}

\subsection{Where do agents spend time and how do they fail (RQ3)?}
\label{sec:error-analysis}

\begin{figure}
    \centering
    \includegraphics[width=\linewidth]{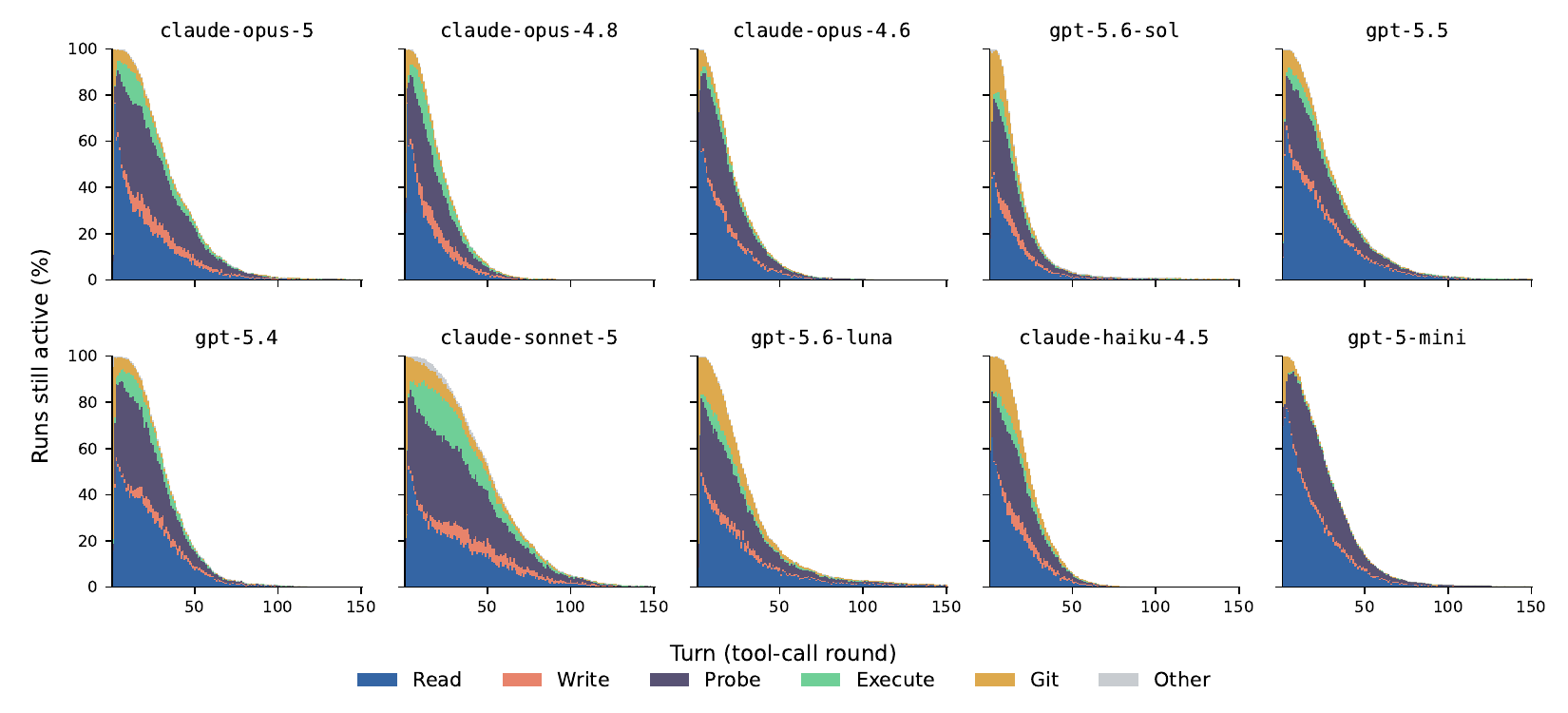}
    \vspace{-15pt}
    \caption{Tool-use composition and run duration across ten models.}
    \label{fig:agent-timeline}
    \vspace{-15pt}
\end{figure}

We found that different agentic models execute similar loops (Figure~\ref{fig:agent-timeline}):
within the loop, agents usually start with git operations (\FirstTurnGit\%, mostly \texttt{git diff} and \texttt{git log}), followed by a long inspection phase in which reading and searching account for \ActionInspect\% of all turn mass (\ActionRead\% Read, \ActionProbe\% Probe), while running code accounts only for \ActionExecute\%.
Agents tend to establish what the change does by looking at source and history far more than by executing it.

Comparing model families, Claude models run code more often than GPT models (\ExecuteLong\% of all turns for \texttt{claude-opus-5} vs. \ExecuteShort\% for \texttt{gpt-5.6-sol}).
Qualitatively, we observe this helps \texttt{claude-opus-5} discover a few bugs that are missed by others:
on a \nolinkurl{yt-dlp} task \texttt{opus-5} retrieved roughly 25 archived pages through the Wayback CDX API, ran the extractor, and reported that livestream pages yield IDs for only two of nine tested pages;
on a GitExtensions task \texttt{opus-5} installed the .NET SDK and built an MSBuild probe to prove a DLL was absent from the publish output.

More noticeably, different models stay in the loop for different length, resulting in different ``effective'' budgets.
Half of the \texttt{gpt-5.6-sol} runs are over by turn \HalfLifeShort{} and only \SurviveShort\%
survive to turn 50, whereas \texttt{claude-opus-5} reaches its half-life at turn \HalfLifeLong{}.
This results in different levels of comprehensiveness of produced review and may explain the result discrepancy partly.

\textit{Failures come from diagnostics instead of retrieval.}
We found \GateInspect\% of agent executions inspect the file with defects, but only \GateReport\% produce a finding for that file, and \GateDiagnose\% actually hit the bugs.
The best-performing model, \texttt{claude-opus-5}, improves much more on producing a finding for the defect file (\BestGateReport\%), but still has gaps in producing a correct diagnosis (\BestGateDiagnose\%).

Qualitatively, we found that many of the failing tasks require knowledge absent from the diff: some require repository-wide reasoning, and some require external context such as library or protocol semantics.
More interestingly, in a few cases, agents are misled by the diff: 
In a Selenium task, all models endorsed the buggy condition because both the in-code comment and the PR description asserted the intended behavior that the code failed to implement. 
In one Elasticsearch task \texttt{opus-5} cited the ground-truth-defective line as its gold standard of correctness and focused on auditing other parts of the same file. 
These cases suggest an overtrust failure: agents accept author claims or existing code as evidence of correctness without independently verifying behavior, even when those descriptions are wrong.
This reflects a larger problem of agents being gullible to its received contexts~\cite{fendley2026comment}, and that \textit{security risks can emerge from automated review processes}.

\mybox{
\faArrowCircleRight~\textbf{Key Findings (RQ3)}:
Review agents run the same read-and-search loop and rarely execute the code they review (\ActionExecute\%). 
They differ more in how long they stay in the loop (from a median of \HalfLifeShort-turn to \HalfLifeLong-turn).
Agents struggle with diagnosis and can be gullible to misleading contexts.
}

\subsection{How do agent design choices affect bug-detection performance (RQ4)?}
\label{sec:rq4-agent-design}

\begin{table*}[t]
\centering
\scriptsize
\renewcommand{\arraystretch}{1.08}
\setlength{\tabcolsep}{2.5pt}
\caption{Comparisons of different prompting strategies for review agent performance.}
\label{tab:prompt-change}
\begin{tabular}{llllllllllllll}
\toprule
\multirow[t]{3}{*}{\textbf{Model}} & \multirow[t]{3}{*}{\textbf{Track}} & \multirow[t]{3}{*}{\textbf{Prompt}} & \multicolumn{4}{l}{\textbf{Recall}} & \multicolumn{2}{l}{\textbf{Review}} & \multicolumn{3}{l}{\textbf{Tokens}} & \textbf{Cost} & \textbf{Latency} \\
\cmidrule(lr){4-7}\cmidrule(lr){8-9}\cmidrule(lr){10-12}
 &  &  & \textbf{@1} & \textbf{@3} & \textbf{@5} & \textbf{@10} & \textbf{Count} & \textbf{Words} & \textbf{In} & \textbf{Cache} & \textbf{Out} & \textbf{(\$)} & \textbf{(min)} \\
 &  &  &  &  &  &  &  & \textbf{(K)} & \textbf{(M)} & \textbf{(M)} & \textbf{(K)} &  &  \\
\midrule
\multirow[t]{4}{*}{\texttt{claude-opus-5}} & \multirow[t]{2}{*}{Cmt.} & Default & 48.3 & 71.2 & 79.6 & 86.0 & 5.76 & 1.08 & 2.43 & 2.34 & 34.3 & 2.55 & 8.6 \\
  & & Comprehensive & 48.3\textsubscript{\textcolor{black}{+0.0}} & 71.4\textsubscript{\textcolor{green!50!black}{+0.2}} & 79.9\textsubscript{\textcolor{green!50!black}{+0.3}} & 87.2\textsubscript{\textcolor{green!50!black}{+1.2}} & 9.26 & 1.66 & 2.41 & 2.31 & 39.6 & 2.76\textsubscript{\textcolor{red!70!black}{+0.2}} & 10.0\textsubscript{\textcolor{red!70!black}{+1.4}} \\
   & \multirow[t]{2}{*}{Fix} & Default & 36.2 & 55.4 & 64.2 & 68.6 & 6.78 & 1.28 & 3.37 & 3.27 & 43.9 & 3.38 & 11.1 \\
  & & Comprehensive & 37.8\textsubscript{\textcolor{green!50!black}{+1.6}} & 53.2\textsubscript{\textcolor{red!70!black}{-2.2}} & 62.4\textsubscript{\textcolor{red!70!black}{-1.8}} & 70.2\textsubscript{\textcolor{green!50!black}{+1.6}} & 10.83 & 1.92 & 3.29 & 3.17 & 50.1 & 3.60\textsubscript{\textcolor{red!70!black}{+0.2}} & 13.0\textsubscript{\textcolor{red!70!black}{+1.9}} \\
\addlinespace[0.2em]
\multirow[t]{4}{*}{\texttt{claude-opus-4.8}} & \multirow[t]{2}{*}{Cmt.} & Default & 45.6 & 65.9 & 70.0 & 71.5 & 2.71 & 0.61 & 1.23 & 1.17 & 24.0 & 1.54 & 6.5 \\
  & & Comprehensive & 47.4\textsubscript{\textcolor{green!50!black}{+1.8}} & 69.5\textsubscript{\textcolor{green!50!black}{+3.5}} & 75.5\textsubscript{\textcolor{green!50!black}{+5.5}} & 78.4\textsubscript{\textcolor{green!50!black}{+6.9}} & 4.40 & 0.81 & 1.16 & 1.09 & 27.1 & 1.67\textsubscript{\textcolor{red!70!black}{+0.1}} & 7.3\textsubscript{\textcolor{red!70!black}{+0.8}} \\
   & \multirow[t]{2}{*}{Fix} & Default & 32.0 & 44.6 & 46.6 & 47.8 & 2.84 & 0.67 & 1.66 & 1.59 & 30.4 & 1.98 & 7.9 \\
  & & Comprehensive & 32.6\textsubscript{\textcolor{green!50!black}{+0.6}} & 47.8\textsubscript{\textcolor{green!50!black}{+3.2}} & 52.6\textsubscript{\textcolor{green!50!black}{+6.0}} & 55.2\textsubscript{\textcolor{green!50!black}{+7.4}} & 5.00 & 0.91 & 1.72 & 1.64 & 36.0 & 2.21\textsubscript{\textcolor{red!70!black}{+0.2}} & 9.2\textsubscript{\textcolor{red!70!black}{+1.3}} \\
\addlinespace[0.2em]
\multirow[t]{4}{*}{\texttt{gpt-5.6-sol}} & \multirow[t]{2}{*}{Cmt.} & Default & 43.5 & 61.9 & 65.6 & 66.6 & 2.72 & 0.34 & 1.01 & 0.86 & 10.4 & 1.50 & 3.7 \\
  & & Comprehensive & 45.3\textsubscript{\textcolor{green!50!black}{+1.8}} & 65.9\textsubscript{\textcolor{green!50!black}{+4.0}} & 71.3\textsubscript{\textcolor{green!50!black}{+5.6}} & 74.2\textsubscript{\textcolor{green!50!black}{+7.6}} & 3.90 & 0.53 & 0.85 & 0.74 & 15.7 & 1.36\textsubscript{\textcolor{green!50!black}{-0.1}} & 5.6\textsubscript{\textcolor{red!70!black}{+1.9}} \\
   & \multirow[t]{2}{*}{Fix} & Default & 29.6 & 45.6 & 49.0 & 51.8 & 3.57 & 0.43 & 1.64 & 1.48 & 16.3 & 2.03 & 5.2 \\
  & & Comprehensive & 30.2\textsubscript{\textcolor{green!50!black}{+0.6}} & 48.8\textsubscript{\textcolor{green!50!black}{+3.2}} & 54.0\textsubscript{\textcolor{green!50!black}{+5.0}} & 57.8\textsubscript{\textcolor{green!50!black}{+6.0}} & 5.39 & 0.71 & 1.16 & 1.03 & 22.1 & 1.81\textsubscript{\textcolor{green!50!black}{-0.2}} & 8.1\textsubscript{\textcolor{red!70!black}{+2.8}} \\
\addlinespace[0.2em]
\multirow[t]{4}{*}{\texttt{gpt-5.5}} & \multirow[t]{2}{*}{Cmt.} & Default & 43.7 & 61.6 & 63.3 & 63.5 & 2.18 & 0.28 & 1.65 & 1.49 & 11.2 & 1.86 & 4.0 \\
  & & Comprehensive & 42.5\textsubscript{\textcolor{red!70!black}{-1.2}} & 62.7\textsubscript{\textcolor{green!50!black}{+1.2}} & 67.4\textsubscript{\textcolor{green!50!black}{+4.2}} & 68.5\textsubscript{\textcolor{green!50!black}{+5.0}} & 3.12 & 0.44 & 1.30 & 1.13 & 14.1 & 1.82\textsubscript{\textcolor{black}{+0.0}} & 4.0\textsubscript{\textcolor{red!70!black}{+0.1}} \\
   & \multirow[t]{2}{*}{Fix} & Default & 29.0 & 45.4 & 47.8 & 48.6 & 2.66 & 0.33 & 2.17 & 2.00 & 15.3 & 2.33 & 5.3 \\
  & & Comprehensive & 30.4\textsubscript{\textcolor{green!50!black}{+1.4}} & 49.0\textsubscript{\textcolor{green!50!black}{+3.6}} & 53.0\textsubscript{\textcolor{green!50!black}{+5.2}} & 55.6\textsubscript{\textcolor{green!50!black}{+7.0}} & 3.95 & 0.54 & 1.75 & 1.63 & 18.4 & 1.97\textsubscript{\textcolor{green!50!black}{-0.4}} & 5.5\textsubscript{\textcolor{red!70!black}{+0.2}} \\
\bottomrule
\end{tabular}
\end{table*}

\begin{table*}[t]
\centering
\scriptsize
\renewcommand{\arraystretch}{1.08}
\setlength{\tabcolsep}{2.5pt}
\caption{Comparisons of different multi-agent configurations for \texttt{gpt-5.6-sol} xhigh.}
\label{tab:multi-agent-change}
\begin{tabular}{lllllllllllll}
\toprule
\multirow[t]{3}{*}{\textbf{Track}} & \multirow[t]{3}{*}{\textbf{Configuration}} & \multicolumn{4}{l}{\textbf{Recall}} & \multicolumn{2}{l}{\textbf{Review}} & \multicolumn{3}{l}{\textbf{Tokens}} & \textbf{Cost} & \textbf{Latency} \\
\cmidrule(lr){3-6}\cmidrule(lr){7-8}\cmidrule(lr){9-11}
 &  & \textbf{@1} & \textbf{@3} & \textbf{@5} & \textbf{@10} & \textbf{Count} & \textbf{Words} & \textbf{In} & \textbf{Cache} & \textbf{Out} & \textbf{(\$)} & \textbf{(min)} \\
 &  &  &  &  &  &  & \textbf{(K)} & \textbf{(M)} & \textbf{(M)} & \textbf{(K)} &  &  \\
\midrule
\multirow[t]{6}{*}{Cmt.} & Base & 43.5 & 61.9 & 65.6 & 66.6 & 2.72 & 0.34 & 1.01 & 0.86 & 10.4 & 1.50 & 3.7 \\
 & Self-refine & 45.3\textsubscript{\textcolor{green!50!black}{+1.8}} & 63.7\textsubscript{\textcolor{green!50!black}{+1.8}} & 67.1\textsubscript{\textcolor{green!50!black}{+1.4}} & 68.2\textsubscript{\textcolor{green!50!black}{+1.6}} & 3.02 & 0.36 & 2.30 & 2.02 & 22.9 & 3.10\textsubscript{\textcolor{red!70!black}{+1.6}} & 9.1\textsubscript{\textcolor{red!70!black}{+5.4}} \\
 & Ensemble@3 & 42.7\textsubscript{\textcolor{red!70!black}{-0.8}} & 64.1\textsubscript{\textcolor{green!50!black}{+2.2}} & 68.0\textsubscript{\textcolor{green!50!black}{+2.4}} & 70.4\textsubscript{\textcolor{green!50!black}{+3.8}} & 3.27 & 0.36 & 3.91 & 3.49 & 42.3 & 5.14\textsubscript{\textcolor{red!70!black}{+3.6}} & 14.9\textsubscript{\textcolor{red!70!black}{+11.2}} \\
 & Ensemble@5 & 44.1\textsubscript{\textcolor{green!50!black}{+0.6}} & 65.1\textsubscript{\textcolor{green!50!black}{+3.2}} & 70.4\textsubscript{\textcolor{green!50!black}{+4.7}} & 72.4\textsubscript{\textcolor{green!50!black}{+5.8}} & 3.41 & 0.38 & 5.93 & 5.22 & 62.9 & 8.07\textsubscript{\textcolor{red!70!black}{+6.6}} & 21.8\textsubscript{\textcolor{red!70!black}{+18.1}} \\
 & Ensemble@3 (diff.\ models) & 45.7\textsubscript{\textcolor{green!50!black}{+2.2}} & 62.8\textsubscript{\textcolor{green!50!black}{+1.0}} & 66.8\textsubscript{\textcolor{green!50!black}{+1.2}} & 68.2\textsubscript{\textcolor{green!50!black}{+1.6}} & 2.88 & 0.33 & 4.09 & 3.63 & 50.4 & 5.57\textsubscript{\textcolor{red!70!black}{+4.1}} & 15.9\textsubscript{\textcolor{red!70!black}{+12.1}} \\
 & Sub-agents & 42.8\textsubscript{\textcolor{red!70!black}{-0.7}} & 58.3\textsubscript{\textcolor{red!70!black}{-3.6}} & 61.9\textsubscript{\textcolor{red!70!black}{-3.7}} & 62.8\textsubscript{\textcolor{red!70!black}{-3.8}} & 2.88 & 0.27 & 3.59 & 3.13 & 46.7 & 5.24\textsubscript{\textcolor{red!70!black}{+3.7}} & 9.7\textsubscript{\textcolor{red!70!black}{+6.0}} \\
\addlinespace[0.2em]
\multirow[t]{6}{*}{Fix} & Base & 29.6 & 45.6 & 49.0 & 51.8 & 3.57 & 0.43 & 1.64 & 1.48 & 16.3 & 2.03 & 5.2 \\
 & Self-refine & 30.8\textsubscript{\textcolor{green!50!black}{+1.2}} & 46.8\textsubscript{\textcolor{green!50!black}{+1.2}} & 52.0\textsubscript{\textcolor{green!50!black}{+3.0}} & 54.8\textsubscript{\textcolor{green!50!black}{+3.0}} & 4.20 & 0.49 & 3.45 & 3.13 & 32.7 & 4.14\textsubscript{\textcolor{red!70!black}{+2.1}} & 12.5\textsubscript{\textcolor{red!70!black}{+7.2}} \\
 & Ensemble@3 & 31.4\textsubscript{\textcolor{green!50!black}{+1.8}} & 50.0\textsubscript{\textcolor{green!50!black}{+4.4}} & 53.6\textsubscript{\textcolor{green!50!black}{+4.6}} & 58.0\textsubscript{\textcolor{green!50!black}{+6.2}} & 4.43 & 0.48 & 6.05 & 5.48 & 63.2 & 7.48\textsubscript{\textcolor{red!70!black}{+5.5}} & 18.6\textsubscript{\textcolor{red!70!black}{+13.3}} \\
 & Ensemble@5 & 30.2\textsubscript{\textcolor{green!50!black}{+0.6}} & 49.4\textsubscript{\textcolor{green!50!black}{+3.8}} & 53.6\textsubscript{\textcolor{green!50!black}{+4.6}} & 57.6\textsubscript{\textcolor{green!50!black}{+5.8}} & 4.81 & 0.52 & 9.37 & 8.50 & 95.8 & 11.47\textsubscript{\textcolor{red!70!black}{+9.4}} & 28.5\textsubscript{\textcolor{red!70!black}{+23.3}} \\
 & Ensemble@3 (diff.\ models) & 32.4\textsubscript{\textcolor{green!50!black}{+2.8}} & 49.6\textsubscript{\textcolor{green!50!black}{+4.0}} & 52.8\textsubscript{\textcolor{green!50!black}{+3.8}} & 56.0\textsubscript{\textcolor{green!50!black}{+4.2}} & 3.84 & 0.44 & 5.85 & 5.41 & 69.5 & 6.91\textsubscript{\textcolor{red!70!black}{+4.9}} & 22.2\textsubscript{\textcolor{red!70!black}{+17.0}} \\
 & Sub-agents & 33.4\textsubscript{\textcolor{green!50!black}{+3.8}} & 50.2\textsubscript{\textcolor{green!50!black}{+4.6}} & 55.0\textsubscript{\textcolor{green!50!black}{+6.0}} & 56.8\textsubscript{\textcolor{green!50!black}{+5.0}} & 4.23 & 0.41 & 5.53 & 4.93 & 71.0 & 7.63\textsubscript{\textcolor{red!70!black}{+5.6}} & 9.7\textsubscript{\textcolor{red!70!black}{+4.4}} \\
\bottomrule
\end{tabular}
\vspace{-10pt}
\end{table*}

First, we found that adjusting effort levels is the simplest yet effective strategy to scale up test-time compute for better performance.
\EffortModel{} can jump from around \EffortLowRecall\% to \EffortMaxRecall\% detection with efforts set from low to max, at \EffortCostRatio x the cost (Figure~\ref{fig:pareto-front}).

Second, we found that prompting agents to output more comprehensive findings is fairly effective, especially for less-verbose models like \texttt{gpt-5.6-sol} (+\PromptGainSol\%), with similar cost (Table~\ref{tab:prompt-change}).
Models do output more findings, but they are also strong in putting important findings towards the top.
In contrast, multi-agent setups provide smaller improvements (+\MultiAgentGainMean\% on average, Table~\ref{tab:multi-agent-change}), with a much higher cost (\MultiAgentCostRatioMean x).
This shows up in the pareto front (Figure~\ref{fig:pareto-front}) that comprehensive prompt configurations dominate multi-agent configurations for test-time compute scaling.

We, however, do notice that improvements from prompt changes vary across models.
Noticeably, for \texttt{claude-opus-5}, the improvements are much smaller or even non-existent compared to earlier models.
This reflects on a constant theme of agent harness design and model capability: many of the encouraged behaviors are gradually incorporated into model training, and hence earlier useful harness designs may become outdated.
It is important to periodically revisit harness designs throughout model evolutions.

For harnesses, we observe \texttt{copilot-cli} to outperform \texttt{claude-code}, justifying our harness choices (Figure~\ref{fig:pareto-front}).
Noticeably, we observed \texttt{claude-opus-5} with \texttt{claude-code} comes with a sharp increase in cost.
Inspecting the trajectories, we found this configuration much more frequently uses sub-agents (\SubagentsBest{} per run), compared to \texttt{copilot-cli} (\SubagentsCopilot{} per run) or \texttt{claude-code} with \texttt{opus-4.8} (\SubagentsPrior{} per run).
This matches Anthropic's official documentation that ``Claude Opus 5 delegates to sub-agents more readily than prior models''~\cite{anthropic_prompting_claude_opus_5}.
Developers need to watch out for such unexpected cost surge with model changes.

\mybox{
\faArrowCircleRight~\textbf{Key Findings (RQ4)}:
Agents benefit from additional test-time compute through effort-level, prompting or multi-agents.
Comprehensive prompting is more cost-effective and performant compared to multi-agents for code review.
}

\section{Discussion}
\label{sec:discussion}

\subsection{When can a software factory be lights-off?}
While coding agents have gained substantial review capabilities, our results suggest that agents are not yet ready for a fully autonomous software factory, especially for large-scale proprietary industry repositories.
Even the strongest agent misses known bugs, with heavy performance drops on proprietary repositories, and failures concentrate in diffs that require non-local or domain-specific reasoning.
The present evidence therefore supports ``lights-off'' operation only for narrowly scoped, low-consequence PRs.
For consequential changes, review agents are better treated as scalable peers that broaden coverage while an accountable human retains the final decision.
While the boundary can move as models improve, developers should calibrate based on repository-specific, continuously monitored evidence rather than aggregate benchmark performance alone.

\subsection{Recommendations for best practices when adopting review agents}
While review agents are not yet capable of fully replacing human reviewers, they are still quite valuable to be incorporated into the review pipeline to help developers catch many bugs they would otherwise miss (Figure~\ref{fig:human-agent}).
The human-in-the-loop adoption, however, should be done with care, based on review contexts.
Based on our empirical observations, we derive the following actionable recommendations for review agent deployment:

\textit{Scale review efforts based on PR complexity.}
An agent's bug-detection capability decreases as a PR becomes larger or contains more kinds of changes (cf. Section~\ref{sec:rq2-conditions}).
Teams should use these properties to allocate both automated and human review effort.
Small, homogeneous changes may be reviewed with a cheaper configuration, whereas large, heterogeneous PRs warrant stronger models, higher reasoning effort, comprehensive prompts, and careful human review.
Splitting unrelated changes into smaller PRs can also reduce the amount of code competing for reviewer attention.
Human reviewers should retain responsibility for project values and long-term maintainability, where our results suggest that their attention is particularly complementary to that of agents.

\textit{Steer review directions based on priorities.}
By default, agent reviews focus on more severe, functional bugs. 
Their review efforts can also be (mis-)directed by surface cues like the types of PR, statements in code comments, or specified behaviors in test cases.
Developers should provide custom instructions that reflect their priorities -- for example, explicitly asking an agent to examine usability, portability, backward compatibility, or maintainability when those properties matter.

\textit{Incorporate repository knowledge for better review quality.}
Agents excel at finding bugs that can be clearly inferred based on local diff contexts.
They struggle more when a correct judgment depends on project-specific knowledge, surrounding source files, or runtime evidence, which is often the case for large, proprietary repositories.
To mitigate these deficiencies, developers can incorporate repository-specific instructions and knowledge bases to guide and improve review quality.

\textit{Monitor review quality and adapt harness over model evolution.}
Review capability changes with both the underlying model and the surrounding environment (e.g., as the codebase evolves).
Harnesses that help one model may provide little benefit to a later model.
Teams should therefore evaluate the deployed agent configuration periodically.
They can apply our data-construction method to construct custom benchmarks for their own repositories, and adapt the harness as needed when model behaviors change.

\subsection{Research opportunities}

\textit{Self-improving review agents.}
One emerging theme from our analysis is that current review agents are still weak at detecting bugs that require project-level knowledge.
While developers can provide explicit scaffold and contextual information, 
future review agents should also be able to maintain memory that enables continual adaptation to (evolving) software repositories.
Such memory can be built from artifacts produced in engineering processes, e.g., design documents, agentic sessions, past review discussions, prior incidents and commits~\cite{wang2026improving, mu2026experepair}.
To build effective review agents, researchers need to create efficient and effective memory representations, such that new information is incorporated and stale information is removed~\cite{lindenbauer2025knowledge}.
They will also need to create effective retrieval mechanisms, and train agents to use the retrieval tools to seek relevant information.
To enable self-improvement, researchers can build on our work, and create reliable data pipelines to mine task instances to guide memory evolution.

\textit{Improve human-in-the-loop review efficiency.}
Our results suggest that human-in-the-loop reviews are still important for bug detection.
However, human reviewers struggle to keep up with agentic code generation~\cite{agarwal20263100}.
To keep humans in the review loop, we need to design effective human-agent collaboration.
Instead of having humans manually review every single line of code, we should build systems to assist and speed up human inspection, and maximize the value they provide at minimal cognitive cost.
For example, researchers can build \textit{code understanding agents}, that provide PR walkthrough, potentially tailored to reviewers' expertise and past interactions.
The agents should be able to highlight the parts that are high-risk, high-impact, or they are least certain about.
Training review agents that are able to calibrate their uncertainty well and communicate it to humans would be a promising direction.

\section{Conclusion}
In this work, we introduce \bench{} and empirically study when and how code review agents detect human-verified bugs.
We show that agents are improving rapidly, but still struggle with complex changes and bugs requiring project context, particularly in proprietary repositories.
Comprehensive prompting improves detection more cost-effectively than multi-agent setups, though its benefits vary across models.
These findings support retaining human oversight and provide guidance for adapting review effort, context, and agent configurations to project needs.

\section*{Data Availability}
We are unable to make the replication package publicly available. 
We provide an open-source version of the benchmark at 
\faGithub~\url{https://github.com/malusamayo/CRBugBench}.

\begin{acks}
We want to thank Kaia Newman, Jina Chun, Jaeyeon Kim, Mike Swafford, Vivien Hung, Selma Ikiz, Karl Piteira, Ben Zorn, Chris Trevino, Nathan Evans, Jonathan Larson, Peter Choi, Rudrajit Choudhuri, and Dax Pryce for their feedback on this work. 
We would also like to thank Jonathan Larson and the Agentic Data Memory team for generously providing the computational resources and infrastructure that made this work possible.
\end{acks}

\bibliographystyle{ACM-Reference-Format}
\bibliography{main}

@String{Computing = "Computing" }

@String{Computer = "{IEEE} Computer" }

@String{Springer = "Springer-Verlag" }

@inproceedings{bacchelli2013expectations,
  title={Expectations, outcomes, and challenges of modern code review},
  author={Bacchelli, Alberto and Bird, Christian},
  booktitle={2013 35th International Conference on Software Engineering (ICSE)},
  pages={712--721},
  year={2013},
  publisher={IEEE},
  doi={10.1109/ICSE.2013.6606617}
}

@article{zeng2025benchmarking,
  title={{SWR-Bench}: Assessing {LLM} Performance in Real-World Code Review Comment Generation},
  author={Zeng, Zhengran and Shi, Ruikai and Han, Keke and Li, Yixin and Sun, Kaicheng and Wang, Yidong and Yu, Zhuohao and Xie, Rui and Ye, Wei and Zhang, Shikun},
  journal={Proceedings of the ACM on Software Engineering},
  volume={3},
  number={FSE},
  articleno={FSE137},
  numpages={23},
  year={2026},
  publisher={Association for Computing Machinery},
  doi={10.1145/3808144}
}

@misc{stackoverflow2025technology,
  author       = {{Stack Overflow}},
  title        = {Technology | 2025 Stack Overflow Developer Survey},
  year         = {2025},
  howpublished = {\url{https://survey.stackoverflow.co/2025/technology/}}
}

@article{sliwerski2005changes,
  title={When do changes induce fixes?},
  author={{\'S}liwerski, Jacek and Zimmermann, Thomas and Zeller, Andreas},
  journal={ACM sigsoft software engineering notes},
  volume={30},
  number={4},
  pages={1--5},
  year={2005},
  publisher={ACM New York, NY, USA}
}

@article{li2025rise,
  title={The rise of ai teammates in software engineering (se) 3.0: How autonomous coding agents are reshaping software engineering},
  author={Li, Hao and Zhang, Haoxiang and Hassan, Ahmed E},
  journal={arXiv preprint arXiv:2507.15003},
  year={2025}
}

@article{bird2022taking,
  title={Taking Flight with Copilot: Early insights and opportunities of AI-powered pair-programming tools},
  author={Bird, Christian and Ford, Denae and Zimmermann, Thomas and Forsgren, Nicole and Kalliamvakou, Eirini and Lowdermilk, Travis and Gazit, Idan},
  journal={Queue},
  volume={20},
  number={6},
  pages={35--57},
  year={2022},
  publisher={ACM New York, NY, USA}
}

@inproceedings{liang2024large,
  title={A large-scale survey on the usability of ai programming assistants: Successes and challenges},
  author={Liang, Jenny T and Yang, Chenyang and Myers, Brad A},
  booktitle={Proceedings of the 46th IEEE/ACM international conference on software engineering},
  pages={1--13},
  year={2024}
}

@misc{zhang2026code,
  title={Code review agent benchmark},
  author={Zhang, Yuntong and Pan, Zhiyuan and Yusuf, Imam Nur Bani and Ruan, Haifeng and Shariffdeen, Ridwan and Roychoudhury, Abhik},
  year={2026},
  eprint={2603.23448},
  archivePrefix={arXiv},
  primaryClass={cs.SE},
  doi={10.48550/arXiv.2603.23448},
  note={Preprint}
}

@inproceedings{li2022automating,
  title={Automating code review activities by large-scale pre-training},
  author={Li, Zhiyu and Lu, Shuai and Guo, Daya and Duan, Nan and Jannu, Shailesh and Jenks, Grant and Majumder, Deep and Green, Jared and Svyatkovskiy, Alexey and Fu, Shengyu and Sundaresan, Neel},
  booktitle={Proceedings of the 30th ACM Joint European Software Engineering Conference and Symposium on the Foundations of Software Engineering},
  pages={1035--1047},
  year={2022},
  publisher={Association for Computing Machinery},
  doi={10.1145/3540250.3549081}
}

@article{fagan1976design,
  author={Fagan, Michael E.},
  title={Design and Code Inspections to Reduce Errors in Program Development},
  journal={IBM Systems Journal},
  volume={15},
  number={3},
  pages={182--211},
  year={1976},
  publisher={IBM},
  doi={10.1147/sj.153.0182}
}

@inproceedings{rigby2013convergent,
  author={Rigby, Peter C. and Bird, Christian},
  title={Convergent Contemporary Software Peer Review Practices},
  booktitle={Proceedings of the 2013 9th Joint Meeting on Foundations of Software Engineering},
  pages={202--212},
  year={2013},
  publisher={Association for Computing Machinery},
  doi={10.1145/2491411.2491444}
}

@inproceedings{mcintosh2014impact,
  author={McIntosh, Shane and Kamei, Yasutaka and Adams, Bram and Hassan, Ahmed E.},
  title={The Impact of Code Review Coverage and Code Review Participation on Software Quality: A Case Study of the {Qt}, {VTK}, and {ITK} Projects},
  booktitle={Proceedings of the 11th Working Conference on Mining Software Repositories},
  pages={192--201},
  year={2014},
  publisher={Association for Computing Machinery},
  doi={10.1145/2597073.2597076}
}

@inproceedings{beller2014problems,
  author={Beller, Moritz and Bacchelli, Alberto and Zaidman, Andy and Juergens, Elmar},
  title={Modern Code Reviews in Open-Source Projects: Which Problems Do They Fix?},
  booktitle={Proceedings of the 11th Working Conference on Mining Software Repositories},
  pages={202--211},
  year={2014},
  publisher={Association for Computing Machinery},
  doi={10.1145/2597073.2597082}
}

@inproceedings{bosu2015characteristics,
  author={Bosu, Amiangshu and Greiler, Michaela and Bird, Christian},
  title={Characteristics of Useful Code Reviews: An Empirical Study at Microsoft},
  booktitle={2015 IEEE/ACM 12th Working Conference on Mining Software Repositories},
  pages={146--156},
  year={2015},
  publisher={IEEE},
  doi={10.1109/MSR.2015.21}
}

@inproceedings{sadowski2018modern,
  author={Sadowski, Caitlin and S{\"o}derberg, Emma and Church, Luke and Sipko, Michal and Bacchelli, Alberto},
  title={Modern Code Review: A Case Study at {Google}},
  booktitle={Proceedings of the 40th International Conference on Software Engineering: Software Engineering in Practice},
  pages={181--190},
  year={2018},
  publisher={Association for Computing Machinery},
  doi={10.1145/3183519.3183525}
}

@article{turzo2024useful,
  author={Turzo, Asif Kamal and Bosu, Amiangshu},
  title={What Makes a Code Review Useful to {OpenDev} Developers? An Empirical Investigation},
  journal={Empirical Software Engineering},
  volume={29},
  number={1},
  articleno={6},
  year={2024},
  publisher={Springer},
  doi={10.1007/s10664-023-10411-x}
}

@inproceedings{balachandran2013reducing,
  author={Balachandran, Vipin},
  title={Reducing Human Effort and Improving Quality in Peer Code Reviews Using Automatic Static Analysis and Reviewer Recommendation},
  booktitle={2013 35th International Conference on Software Engineering (ICSE)},
  pages={931--940},
  year={2013},
  publisher={IEEE},
  doi={10.1109/ICSE.2013.6606642}
}

@inproceedings{tufano2021towards,
  author={Tufano, Rosalia and Pascarella, Luca and Tufano, Michele and Poshyvanyk, Denys and Bavota, Gabriele},
  title={Towards Automating Code Review Activities},
  booktitle={2021 IEEE/ACM 43rd International Conference on Software Engineering (ICSE)},
  pages={163--174},
  year={2021},
  publisher={IEEE},
  doi={10.1109/ICSE43902.2021.00027}
}

@inproceedings{tufano2022using,
  author={Tufano, Rosalia and Masiero, Simone and Mastropaolo, Antonio and Pascarella, Luca and Poshyvanyk, Denys and Bavota, Gabriele},
  title={Using Pre-Trained Models to Boost Code Review Automation},
  booktitle={Proceedings of the 44th International Conference on Software Engineering},
  pages={2291--2302},
  year={2022},
  publisher={Association for Computing Machinery},
  doi={10.1145/3510003.3510621}
}

@inproceedings{naik2025crscore,
  author={Naik, Atharva and Alenius, Marcus and Fried, Daniel and Rose, Carolyn},
  title={{CRScore}: Grounding Automated Evaluation of Code Review Comments in Code Claims and Smells},
  booktitle={Proceedings of the 2025 Conference of the Nations of the Americas Chapter of the Association for Computational Linguistics: Human Language Technologies (Volume 1: Long Papers)},
  pages={9049--9076},
  year={2025},
  publisher={Association for Computational Linguistics},
  doi={10.18653/v1/2025.naacl-long.457}
}

@inproceedings{tufano2025doubleedged,
  author={Tufano, Rosalia and Martin-Lopez, Alberto and Tayeb, Ahmad and Dabi{\'c}, Ozren and Haiduc, Sonia and Bavota, Gabriele},
  title={Deep Learning-Based Code Reviews: A Paradigm Shift or a Double-Edged Sword?},
  booktitle={2025 IEEE/ACM 47th International Conference on Software Engineering (ICSE)},
  pages={1640--1652},
  year={2025},
  publisher={IEEE},
  doi={10.1109/ICSE55347.2025.00060}
}

@inproceedings{cihan2025automated,
  author={Cihan, Umut and Haratian, Vahid and {\.I}{\c c}{\"o}z, Arda and G{\"u}l, Mert Kaan and Devran, {\"O}mercan and Bayendur, Emircan Furkan and U{\c c}ar, Baykal Mehmet and T{\"u}z{\"u}n, Eray},
  title={Automated Code Review in Practice},
  booktitle={2025 IEEE/ACM 47th International Conference on Software Engineering: Software Engineering in Practice (ICSE-SEIP)},
  pages={425--436},
  year={2025},
  publisher={IEEE},
  doi={10.1109/ICSE-SEIP66354.2025.00043}
}

@article{yang2026roadmap,
  author={Yang, Zezhou and Gao, Cuiyun and Guo, Zhaoqiang and Li, Zhenhao and Liu, Kui and Xia, Xin and Zhou, Yuming},
  title={A Roadmap for Modern Code Review: Challenges and Opportunities},
  journal={ACM Transactions on Software Engineering and Methodology},
  year={2026},
  publisher={Association for Computing Machinery},
  doi={10.1145/3800963},
  note={Just Accepted}
}

@inproceedings{lu2025deepcrceval,
  author={Lu, Junyi and Li, Xiaojia and Hua, Zihan and Yu, Lei and Cheng, Shiqi and Yang, Li and Zhang, Fengjun and Zuo, Chun},
  title={{DeepCRCEval}: Revisiting the Evaluation of Code Review Comment Generation},
  booktitle={Fundamental Approaches to Software Engineering},
  series={Lecture Notes in Computer Science},
  volume={15693},
  pages={43--64},
  year={2025},
  publisher={Springer},
  doi={10.1007/978-3-031-90900-9_3}
}

@inproceedings{hu2026contextcrbench,
  author={Hu, Ruida and Wang, Xinchen and Wen, Xin-Cheng and Zhang, Zhao and Jiang, Bo and Gao, Pengfei and Peng, Chao and Gao, Cuiyun},
  title={Benchmarking {LLM}s for Fine-Grained Code Review with Enriched Context in Practice},
  booktitle={Proceedings of the 34th ACM International Conference on the Foundations of Software Engineering, FSE Companion 2026},
  pages={620--631},
  year={2026},
  publisher={Association for Computing Machinery},
  doi={10.1145/3803437.3805235}
}

@inproceedings{tang2024codeagent,
  author={Tang, Xunzhu and Kim, Kisub and Song, Yewei and Lothritz, Cedric and Li, Bei and Ezzini, Saad and Tian, Haoye and Klein, Jacques and Bissyand{\'e}, Tegawend{\'e} F.},
  title={{CodeAgent}: Autonomous Communicative Agents for Code Review},
  booktitle={Proceedings of the 2024 Conference on Empirical Methods in Natural Language Processing},
  pages={11279--11313},
  year={2024},
  publisher={Association for Computational Linguistics},
  doi={10.18653/v1/2024.emnlp-main.632}
}

@inproceedings{tantithamthavorn2026rovodev,
  author={Tantithamthavorn, Chakkrit Kla and Zou, Yaotian and Wong, Andy and Gupta, Michael and Wang, Zhe and Buller, Mike and Jiang, Ryan and Watson, Matthew and Jeong, Minwoo and Chen, Kun and Wu, Ming},
  title={{RovoDev} Code Reviewer: A Large-Scale Online Evaluation of {LLM}-Based Code Review Automation at Atlassian},
  booktitle={Proceedings of the 2026 IEEE/ACM 48th International Conference on Software Engineering: Software Engineering in Practice},
  pages={34--45},
  year={2026},
  publisher={Association for Computing Machinery},
  doi={10.1145/3786583.3786851}
}

@online{shapiro2026fivelevels,
  author  = {Dan Shapiro},
  title   = {The Five Levels: From Spicy Autocomplete to the Dark Factory},
  date    = {2026-01-23},
  url     = {https://www.danshapiro.com/blog/2026/01/the-five-levels-from-spicy-autocomplete-to-the-software-factory/},
  urldate = {2026-08-05}
}

@online{anthropicClaudeCode,
  author  = {{Anthropic}},
  title   = {Claude Code},
  year = {2025},
  url     = {https://claude.com/product/claude-code}
}

@online{anthropicCodeReview,
  author  = {{Anthropic}},
  title   = {Code Review},
  year = {2026},
  url     = {https://code.claude.com/docs/en/code-review}
}

@article{pereira2026cr,
  title={{CR-Bench}: Evaluating the Real-World Utility of {AI} Code Review Agents},
  author={Pereira, Kristen and Sinha, Neelabh and Ghosh, Rajat and Dutta, Debojyoti},
  journal={arXiv preprint arXiv:2603.11078},
  year={2026},
  doi={10.48550/arXiv.2603.11078}
}

@article{agarwal20263100,
  title={3100 Opinions on Code Review in an AI World: Building Causal Theory from Practitioner Discourse},
  author={Agarwal, Shyam and Miller, Courtney and K{\"a}stner, Christian and Vasilescu, Bogdan},
  journal={arXiv preprint arXiv:2607.07980},
  year={2026}
}

@inproceedings{bird2011ownership,
  author={Bird, Christian and Nagappan, Nachiappan and Murphy, Brendan and Gall, Harald and Devanbu, Premkumar},
  title={Don't Touch My Code! Examining the Effects of Ownership on Software Quality},
  booktitle={Proceedings of the 19th ACM SIGSOFT Symposium and the 13th European Conference on Foundations of Software Engineering},
  pages={4--14},
  year={2011},
  publisher={Association for Computing Machinery},
  doi={10.1145/2025113.2025119}
}

@inproceedings{dambros2010comparison,
  author={D'Ambros, Marco and Lanza, Michele and Robbes, Romain},
  title={An Extensive Comparison of Bug Prediction Approaches},
  booktitle={2010 7th IEEE Working Conference on Mining Software Repositories},
  pages={31--41},
  year={2010},
  publisher={IEEE},
  doi={10.1109/MSR.2010.5463279}
}

@inproceedings{lu2025practical,
  author={Lu, Junyi and Jiang, Lili and Li, Xiaojia and Fang, Jianbing and Zhang, Fengjun and Yang, Li and Zuo, Chun},
  title={Towards Practical Defect-Focused Automated Code Review},
  booktitle={Proceedings of the 42nd International Conference on Machine Learning},
  pages={40505--40536},
  volume={267},
  series={Proceedings of Machine Learning Research},
  year={2025},
  publisher={PMLR},
  url={https://proceedings.mlr.press/v267/lu25f.html}
}

@article{zheng2023judging,
  title={Judging llm-as-a-judge with mt-bench and chatbot arena},
  author={Zheng, Lianmin and Chiang, Wei-Lin and Sheng, Ying and Zhuang, Siyuan and Wu, Zhanghao and Zhuang, Yonghao and Lin, Zi and Li, Zhuohan and Li, Dacheng and Xing, Eric and others},
  journal={Advances in neural information processing systems},
  volume={36},
  pages={46595--46623},
  year={2023}
}

@techreport{iso25010_2011,
  author      = {{ISO/IEC}},
  title       = {Systems and software engineering --- Systems and software
                 Quality Requirements and Evaluation ({SQuaRE}) ---
                 System and software quality models},
  institution = {International Organization for Standardization and
                 International Electrotechnical Commission},
  type        = {International Standard},
  number      = {ISO/IEC 25010:2011},
  year        = {2011},
  url         = {https://www.iso.org/standard/35733.html}
}

@article{benjamini1995controlling,
  author  = {Benjamini, Yoav and Hochberg, Yosef},
  title   = {Controlling the False Discovery Rate: A Practical and Powerful Approach to Multiple Testing},
  journal = {Journal of the Royal Statistical Society: Series B (Methodological)},
  year    = {1995},
  volume  = {57},
  number  = {1},
  pages   = {289--300},
  doi     = {10.1111/j.2517-6161.1995.tb02031.x}
}

@article{fendley2026comment,
  title={Comment and Control: Hijacking Agentic Workflows via Context-Grounded Evolution},
  author={Fendley, Neil and Liu, Zhengyu and Guan, Aonan and Zhong, Jiacheng and Cao, Yinzhi},
  journal={arXiv preprint arXiv:2605.11229},
  year={2026}
}

@misc{anthropic_prompting_claude_opus_5,
  author       = {{Anthropic}},
  title        = {Prompting {Claude Opus 5}},
  howpublished = {Claude Platform Docs},
  url          = {https://platform.claude.com/docs/en/build-with-claude/prompt-engineering/prompting-claude-opus-5},
}

@inproceedings{wang2026improving,
  title={Improving code localization with repository memory},
  author={Wang, Boshi and Xu, Weijian and Li, Yunsheng and Gao, Xuemei and Xie, Yujia and Sun, Huan and Chen, Dongdong},
  booktitle={International Conference on Learning Representations},
  volume={2026},
  pages={111266--111285},
  year={2026}
}

@article{mu2026experepair,
  title={Experepair: Dual-memory enhanced llm-based repository-level program repair},
  author={Mu, Fangwen and Wang, Junjie and Shi, Lin and Wang, Song and Li, Shoubin and Wang, Qing},
  journal={Proceedings of the ACM on Software Engineering},
  volume={3},
  number={FSE},
  pages={3954--3976},
  year={2026},
  publisher={ACM New York, NY, USA}
}

@inproceedings{lindenbauer2025knowledge,
  title={From knowledge to noise: CTIM-rover and the pitfalls of episodic memory in software engineering agents},
  author={Lindenbauer, Tobias and Groh, Georg and Sch{\"u}tze, Hinrich},
  booktitle={Proceedings of the 1st Workshop for Research on Agent Language Models (REALM 2025)},
  pages={411--427},
  year={2025}
}

@article{sarma2024effective,
  title={Effective teaching through code reviews: Patterns and anti-patterns},
  author={Sarma, Anita and Chen, Nina},
  journal={Proceedings of the ACM on Software Engineering},
  volume={1},
  number={FSE},
  pages={1262--1283},
  year={2024},
  publisher={ACM New York, NY, USA}
}

@article{baum2019associating,
  author  = {Tobias Baum and Kurt Schneider and Alberto Bacchelli},
  title   = {Associating Working Memory Capacity and Code Change Ordering with Code Review Performance},
  journal = {Empirical Software Engineering},
  volume  = {24},
  number  = {4},
  pages   = {1762--1798},
  year    = {2019},
  doi     = {10.1007/s10664-018-9676-8}
}

@inproceedings{ram2018reviewability,
  author    = {Achyudh Ram and Anand Ashok Sawant and Marco Castelluccio and Alberto Bacchelli},
  title     = {What Makes a Code Change Easier to Review: An Empirical Investigation on Code Change Reviewability},
  booktitle = {Proc. 26th ACM Joint European Software Engineering Conference and Symposium on the Foundations of Software Engineering (ESEC/FSE)},
  year      = {2018},
  pages     = {201--212}
}

@inproceedings{tao2012understand,
  author    = {Yida Tao and Yingnong Dang and Tao Xie and Dongmei Zhang and Sunghun Kim},
  title     = {How Do Software Engineers Understand Code Changes? An Exploratory Study in Industry},
  booktitle = {Proc. 20th ACM SIGSOFT International Symposium on the Foundations of Software Engineering (FSE)},
  year      = {2012},
  pages     = {51:1--51:11},
  doi       = {10.1145/2393596.2393656}
}

@inproceedings{barnett2015decomposition,
  author    = {Mike Barnett and Christian Bird and Jo{\~a}o Brunet and Shuvendu K. Lahiri},
  title     = {Helping Developers Help Themselves: Automatic Decomposition of Code Review Changesets},
  booktitle = {Proc. 37th International Conference on Software Engineering (ICSE)},
  year      = {2015},
  pages     = {134--144},
  doi       = {10.1109/ICSE.2015.35}
}

@inproceedings{spadini2018testing,
  author    = {Davide Spadini and Maur{\'\i}cio Aniche and Margaret-Anne Storey and Magiel Bruntink and Alberto Bacchelli},
  title     = {When Testing Meets Code Review: Why and How Developers Review Tests},
  booktitle = {Proc. 40th International Conference on Software Engineering (ICSE)},
  year      = {2018},
  pages     = {677--687},
  doi       = {10.1145/3180155.3180192}
}

@inproceedings{dunsmore2000delocalisation,
  author    = {Alastair Dunsmore and Marc Roper and Murray Wood},
  title     = {Object-Oriented Inspection in the Face of Delocalisation},
  booktitle = {Proc. 22nd International Conference on Software Engineering (ICSE)},
  year      = {2000},
  pages     = {467--476},
  doi       = {10.1145/337180.337343}
}

@article{mantyla2009types,
  author  = {Mika V. M{\"a}ntyl{\"a} and Casper Lassenius},
  title   = {What Types of Defects Are Really Discovered in Code Reviews?},
  journal = {IEEE Transactions on Software Engineering},
  volume  = {35},
  number  = {3},
  pages   = {430--448},
  year    = {2009},
  doi     = {10.1109/TSE.2008.71}
}

@article{zhang2026aacr,
  title={AACR-Bench: Evaluating Automatic Code Review with Holistic Repository-Level Context},
  author={Zhang, Lei and Yu, Yongda and Yu, Minghui and Guo, Xinxin and Zhuang, Zhengqi and Rong, Guoping and Shao, Dong and Shen, Haifeng and Kuang, Hongyu and Li, Zhengfeng and others},
  journal={arXiv preprint arXiv:2601.19494},
  year={2026}
}

\end{document}